\documentclass[aps,prd,reprint,showkeys,noshowpacs,longbibliography,floatfix,amsmath,amssymb,nofootinbib,noshowpacs,superscriptaddress,notitlepage,preprintnumbers]{revtex4-2}

\usepackage[utf8]{inputenc}
\usepackage[english]{babel}
\usepackage[normalem]{ulem}
\usepackage{graphicx}
\usepackage{mathtools}
\usepackage{MnSymbol}
\usepackage{multirow}
\usepackage{tabularx}
\newcolumntype{C}[1]{>{\hsize=#1\hsize\centering\arraybackslash}X}
\newcolumntype{R}[1]{>{\hsize=#1\hsize\raggedleft\arraybackslash}X}
\newcolumntype{L}[1]{>{\hsize=#1\hsize\raggedright\arraybackslash}X}

\usepackage{xr-hyper}
\usepackage[pdftex,bookmarks=true]{hyperref}
\hypersetup{
   colorlinks,
   citecolor=blue,
   filecolor=black,
   linkcolor=blue,
   urlcolor=blue
}
\usepackage{cleveref}
\Crefformat{equation}{Eq.~(#2#1#3)}
\Crefmultiformat{equation}{Eqs.~(#2#1#3)}{ and~(#2#1#3)}{, (#2#1#3)}{ and~(#2#1#3)}
\Crefformat{figure}{Fig.~#2#1#3}
\Crefformat{section}{Sec.~#2#1#3}

\allowdisplaybreaks

\def\pdotq{p \cdot q}

\newcommand{\bv}[1]{{\bf{#1}}}
\newcommand{\bs}[1]{{\boldsymbol{#1}}}
\newcommand{\kb}{\bv{k}}
\newcommand{\eb}{\bv{e}}
\newcommand{\qb}{\bv{q}}
\newcommand{\pb}{\bv{p}}

\def\bslam{\bs{\lambda}}

\def\latmom{\frac{2\pi}{L}}

\begin{document}

\title{Gross-Llewellyn Smith sum rule from lattice QCD}
\author{K.~U.~Can}
\affiliation{CSSM, Department of Physics, The University of Adelaide, Adelaide SA 5005, Australia}
\author{J.~A.~Crawford}
\affiliation{CSSM, Department of Physics, The University of Adelaide, Adelaide SA 5005, Australia}
\author{R.~Horsley}
\affiliation{School of Physics and Astronomy, University of Edinburgh, Edinburgh EH9 3JZ, UK}
\author{P.~E.~L.~Rakow}
\affiliation{Theoretical Physics Division, Department of Mathematical Sciences, University of Liverpool, Liverpool L69 3BX, UK}
\author{T.~G.~Schar}
\affiliation{CSSM, Department of Physics, The University of Adelaide, Adelaide SA 5005, Australia}
\author{G.~Schierholz}
\affiliation{Deutsches Elektronen-Synchrotron DESY, Notkestr. 85, 22607 Hamburg, Germany.}
\author{H.~St\"{u}ben}
\affiliation{Regionales Rechenzentrum, Universit\"{a}t Hamburg, 20146 Hamburg, Germany}
\author{R.~D.~Young}
\affiliation{CSSM, Department of Physics, The University of Adelaide, Adelaide SA 5005, Australia}
\author{J.~M.~Zanotti}
\affiliation{CSSM, Department of Physics, The University of Adelaide, Adelaide SA 5005, Australia}

\collaboration{QCDSF Collaboration}
\noaffiliation
\begin{abstract}
	We compute the Gross-Llewellyn Smith sum rule, i.e. the lowest odd moment of the parity-violating structure function, $F_3$, of the nucleon from a lattice QCD calculation of the Compton amplitude. Our calculations are performed on $48^3 \times 96$ lattices at the $SU(3)$ symmetric point for two lattice spacings. We extract the moments for several values of the current momenta in the range $0.5 \lesssim Q^2 \lesssim 10 \; {\rm GeV}^2$, covering both the nonperturbative and perturbative regimes. We compare our moments to the Gross-Llewellyn Smith sum rule and discuss the implications for higher-twist effects, a determination of $\alpha_s(Q^2)$ from a hadronic quantity complementing the phenomenological and other lattice approaches, and electroweak box contributions crucial for Cabibbo–Kobayashi–Maskawa matrix unitarity studies. 
\end{abstract}
\date{\today}
\maketitle

\section{Introduction}
The parity-odd structure function, $F_3$, accessible via neutrino deep-inelastic scattering ($\nu$DIS), plays a significant role in understanding processes involving quantum chromodynamics (QCD) and electroweak sectors. Its first moment defines the Gross-Llewellyn Smith (GLS) sum rule~\cite{Gross:1969jf}
\begin{equation} \label{eq:GLS}
	I_{\rm GLS}(Q^2) = \frac{1}{2} \int_0^1 dx [F_3^{\nu p} + F_3^{\bar\nu p}](x,Q^2), 
\end{equation}
which in the quark-parton model counts the number of valence quarks in a proton and the perturbatively calculable radiative corrections provide a significant test of the validity of QCD as the theory of strong interactions. While the charged-current interaction, e.g. $W$-exchange, is directly proportional to the number of valence quarks, the neutral-current interaction, e.g. the $\gamma Z$ interference, has an analogous sum rule where the number of valence quarks are weighted by electroweak charges of the quarks, therefore carrying equivalent information as the GLS sum rule. 

Thanks to $F_3$ being a non-singlet quantity, the GLS sum rule is theoretically clean, i.e. it has a vanishing anomalous dimension at leading-order in QCD~\cite{Retey:2000nq}. 
The GLS sum rule is unique in the sense that the nucleon matrix element of leading twist, which is the vector charge, is known analytically and the Wilson coefficient is known to next-to-next-to-next-to-leading order (N${}^3$LO)~\cite{Larin:1991tj,Hayen:2020cxh}, which makes it the ideal laboratory for studying higher twist~\cite{Mueller:1993pa}. 
This also makes the GLS sum rule, along with the Bjorken sum rule, an ideal laboratory for determining $\alpha_s$~\cite{Kim:1998kia,Mangano:2001mj} at the brink of the nonperturbative regime and also for a nonperturbative definition of the strong coupling via an effective charge approach analogous to QED (see~\cite{Deur:2016tte,Deur:2023dzc} for reviews). A first-principles calculation is highly desirable to provide insights into the interplay between the perturbative and nonperturbative regimes, and an additional determination of the strong coupling based on a hadronic observable.

Beyond the GLS sum rule, the $F_3$ structure function holds vital importance for electroweak radiative corrections, which remain one of the main sources of theoretical uncertainty in the determination of the Cabibbo–Kobayashi–Maskawa (CKM) matrix elements and the weak mixing angle at low scales~\cite{Hardy:2014qxa,Hardy:2020qwl,Kumar:2013yoa,Gorchtein:2023naa}. These radiative corrections are directly proportional to the contribution of the electroweak box diagrams, e.g. the $\gamma Z$, or $\gamma W^\pm$ interference, which are sensitive to processes involving hadronic scales. Modern dispersive treatments of the box diagrams~\cite{Blunden:2011rd,PhysRevLett.96.032002,PhysRevLett.121.241804,Seng:2018qru,Shiells:2020fqp} (and see~\cite{Arrington2011782,Gorchtein:2023naa} for reviews) provide a connection to the (first moment of) parity-odd structure function, hence require experimental input. However, the coverage of available experimental data is rather poor, where it either does not cover the low-$Q^2$ region relevant for the hadronic scales or is available for a different isospin channel which would require modelling to relate it to the necessary structure function. In light of this, a nonperturbative approach comes as a powerful tool to determine the box diagram contributions and recent lattice QCD calculations~\cite{Feng:2020zdc,Ma:2021azh,Ma:2023kfr,Yoo:2022lmt,Yoo:2023gln} have indeed provided valuable input to studying CKM unitarity. 

The Feynman-Hellmann approach has proven to be an effective technique to extract the Compton amplitude in lattice QCD~\cite{PhysRevLett.118.242001}, with applications to the calculations of the moments of unpolarised~\cite{PhysRevD.102.114505,QCDSFUKQCDCSSM:2022ncb} and polarised~\cite{Can:2022chd} parity-conserving nucleon structure functions, and generalised parton distributions~\cite{Alec:2021lkf,Hannaford-Gunn:2024aix}. Building upon previous work, we compute the GLS sum rule by calculating the first moment of the $\gamma Z$ interference structure function, $F_3^{\gamma Z}$.

The rest of this paper is organised as follows: In \Cref{sec:ca} we show the Lorentz decomposition of the Compton amplitude and the kinematics we choose to isolate the parity-odd Compton amplitude. The Feynman-Hellmann approach employed to calculate the Compton amplitude is explained in \Cref{sec:fh}, followed by the details of our lattice calculations and analyses in \Cref{sec:simu}. We present and discuss our results \Cref{sec:res} before providing a summary and our concluding remarks in \Cref{sec:sum}. Some further details of our calculations are collected in \Cref{app:lpt,app:f3_el}.
\section{Parity-odd Compton amplitude} \label{sec:ca}
The starting point of the calculation is the time-ordered product of vector and axial-vector currents sandwiched between nucleon states, forming the Compton tensor,
\begin{equation} \label{eq:ctensor}
  T_{\mu\nu}(p,q) = i \rho_{s s^\prime} \hspace{-1.5mm} \int \hspace{-1mm} d^4z e^{i q \cdot z} \langle p,s^\prime | \mathcal{T}\{J_\mu(z) J_\nu^A(0)\} | p,s \rangle,
\end{equation} 
where $p$ ($q$) is the nucleon (current) momentum, and $\rho_{s s^\prime}$ is the spin density matrix with the vector and axial-vector currents defined as,
\begin{align}\label{eq:currents}
	J_\mu(x) &= \sum_{f=u,d} \mathcal{Q}_{f} \bar \psi_f(x) \gamma_\mu \psi_f(x), \\
	J^A_\nu(y) &= \sum_{f=u,d} g_A^{f} \bar \psi_f(y) \gamma_\mu \gamma_5 \psi_f(y),
\end{align}
where $\mathcal{Q}_f$ is the electric charge of the quark of flavour $f$, and $g_A^f = \pm 1/2$ is the axial charge for an up- and down-type quark respectively.

The most general Lorentz decomposition of the Compton tensor in Minkowski space for the spin-averaged case is~\cite{Ji:1993ey,Thomas:2001kw},
\begin{align} \label{eq:f3}
		T_{\mu\nu}(p,q) =& -g_{\mu\nu} \mathcal{F}_1(\omega,Q^2) 
		+ \frac{p_\mu p_\nu}{\pdotq} \mathcal{F}_2(\omega,Q^2) \nonumber \\
		&+ i \, \varepsilon^{\mu\nu\alpha\beta} \frac{p_\alpha q_\beta}{2 \pdotq}\mathcal{F}_3(\omega,Q^2)
		+ \frac{q_\mu q_\nu}{\pdotq} \mathcal{F}_4(\omega,Q^2) \nonumber \\
		&+ \frac{p_{{}_{\{}\mu} q_{\nu_{\}}}}{\pdotq} \mathcal{F}_5(\omega,Q^2)
		+ \frac{p_{{}_{[}\mu} q_{\nu_{]}}}{\pdotq} \mathcal{F}_6(\omega,Q^2),
\end{align}
where $\varepsilon_{0123} = 1$, $\omega = 2 p \cdot q / Q^2$, $Q^2=-q^2$, and $p_{{}_{\{}\mu} q_{\nu_{\}}} = (p_\mu q_\nu + p_\nu q_\mu)/2$ and $p_{{}_{[}\mu} q_{\nu_{]}} = (p_\mu q_\nu - p_\nu q_\mu)/2$ denote the symmetrisation and anti-symmetrisation of the indices. We focus on the $\mathcal{F}_3$ term since the rest can be eliminated for certain choices of kinematics. The analyticity of the Compton tensor along with crossing symmetry and the optical theorem, $\operatorname{Im} \mathcal{F}_3(\omega,Q^2) = 2 \pi F_3(x=1/\omega,Q^2)$, allows a connection between the Compton structure function $\mathcal{F}_3$ and the unpolarised parity-violating physical structure function $F_3$ through the dispersion relation,
\begin{equation} \label{eq:disp}
  \mathcal{F}_3(\omega, Q^2) = 4 \omega \int_0^1 dx \frac{F_3(x,Q^2)}{1-x^2 \omega^2}.
\end{equation}
Upon expanding the geometric series, $(1-x^2 \omega^2)^{-1}$, we arrive at,
\begin{equation} \label{eq:ca_expand}
  \frac{\mathcal{F}_3(\omega, Q^2)}{\omega} = 4 \sum_{n=1,2,\dots} \omega^{2n-2} \, M_{2n-1}^{(3)}(Q^2),
\end{equation}
with the odd Mellin (Cornwall-Norton) moments of $F_3$ defined as,
\begin{equation} \label{eq:moments}
  M_{2n-1}^{(3)}(Q^2) = \int_0^1 dx \, x^{2n-2} \, F_3(x,Q^2),
\end{equation}
for $n=1,2,3,\dots$. Previous studies that extract the lowest even Mellin moment of $F_1$ required a polynomial fit in $\omega$~\cite{PhysRevD.102.114505}, however, the lowest Mellin moment of $F_3$, $M_1^{(3)}(Q^2)$, is directly accessible at $\omega=0$, i.e. for a nucleon at rest, thus can be determined directly from the Compton amplitude without a polynomial fit in $\omega$, noting that the RHS of \Cref{eq:ca_expand} is well defined at $\omega=0$.

\section{Feynman-Hellmann approach} \label{sec:fh}
A computation of the Compton amplitude requires the evaluation of a matrix element that involves two currents, which is a four-point correlation function. However, this is a challenging calculation to undertake on the lattice because of the rapid deterioration of the signal for large time separations and the contamination due to excited states that are harder to control compared to two- or three-point correlators. The application of the Feynman-Hellmann theorem, on the other hand, reduces the problem to a simpler analysis of two-point correlation functions using the established techniques of spectroscopy. 

The Compton tensor, \Cref{eq:ctensor}, involves the product of a vector and an axial-vector current, which requires a generalisation of the Feynman-Hellmann technique presented in~\cite{PhysRevD.102.114505} to include mixed currents. We follow the generalisation applied in \cite{Alec:2021lkf}. It is clear from \Cref{eq:f3} that the $\mathcal{F}_3$ Compton structure function lives in the antisymmetric part of the tensor. In this case, we introduce two spatially oscillating background fields to the fermion action
\begin{align}\label{eq:fh_action}
	\begin{split}
		S(\bslam, q) = S_0 
		  &+ \lambda_1 \int d^4z  (e^{i q \cdot z} + e^{-i q \cdot z}) \mathcal{J}_\mu(z) \\
		  &-i \lambda_2 \int d^4y (e^{i q \cdot y} - e^{-i q \cdot y}) \mathcal{J}^A_\nu(y),
	\end{split}
\end{align}
where $S_0$ is the unperturbed action, and $\bslam \equiv (\lambda_1,\lambda_2)$ denote the strengths of the couplings between the quarks and the external fields. The currents $\mathcal{J}_\mu(z) = Z_V \bar{\psi}_f(z) \gamma_\mu \psi_f(z)$ and $\mathcal{J}_\nu^A(y) = Z_A \bar{\psi}_f(y) \gamma_5 \gamma_\nu \psi_f(y)$ are the renormalised vector and axial vector currents coupling to the quarks of flavour $f$ along the $\mu$ and $\nu$ directions. Here, $q=(\qb,0)$ is the external momentum inserted by the currents and $q_4=0$ by construction in the Feynman-Hellmann approach~\cite{PhysRevD.102.114505}. $Z_{V,A}$ are the renormalisation constants for the local vector and axial vector currents. An advantage of our approach is that the renormalisation is trivial, i.e., we only need a multiplicative renormalisation with known constants $Z_{V,A}$~\cite{Constantinou:2014fka}. The perturbation is introduced on the valence quarks only, hence only quark-line connected contributions are taken into account in this work. The perturbation of sea quarks has been addressed in~\cite{Chambers2015} in the context of disconnected diagram contributions to the nucleon spin.

Following the derivation presented in~\cite{PhysRevD.102.114505}, we arrive at the Feynman-Hellmann relation between the second-order energy shift and the Compton amplitude, 
\begin{equation} \label{eq:secondorder_fh}
    \left. \frac{\partial^2 E_{N_\lambda}(\pb, \qb)}{\partial \lambda_1 \partial \lambda_2} \right|_{\bslam=0} 
    = i \frac{T_{\mu\nu}(p,q) - T_{\mu\nu}(p,-q)}{2 E_{N}(\pb)},
\end{equation}
where $T_{\mu\nu}$ is defined in \Cref{eq:ctensor}, and $E_{N_\lambda}(\pb, \qb)$ is the perturbed nucleon energy with momentum $\pb$ and depends on the current insertion of momentum $\qb$. The $i$ factor and the minus sign in \Cref{eq:secondorder_fh} are due to the second perturbation term added in \Cref{eq:fh_action}. We note that \Cref{eq:secondorder_fh} renders a relation between the second-order energy shift and the Compton amplitude, and in order to isolate the $\mathcal{F}_3$ Compton structure function we make a certain choice of kinematics (discussed in \Cref{sec:simu}). An independent derivation of the relation between the energy shift and the Compton amplitude, based on an expansion of the Lagrangian in terms of a periodic external source~\cite{Agadjanov:2016cjc}, can be found in~\cite{Seng:2019plg}.

\subsection{Flavour decomposition and extracting the energy shifts}
The Feynman-Hellmann implementation we have briefly described above corresponds to the insertion of an external current to a quark line when the quark propagator is computed with the perturbed action, \Cref{eq:fh_action}. Therefore the flavour decomposition is achieved by computing the propagator with the perturbed action, while the action of the spectator quark is left unmodified. When both currents are inserted onto the same quark, we evaluate the flavour-diagonal $uu$ or $dd$ contributions to the Compton amplitude. It is also possible to isolate the mixed-flavour higher-twist $ud$ piece by using positive and negative $\lambda$ pairs~\cite{Hannaford-Gunn:2020pvu,QCDSFUKQCDCSSM:2022ncb}. We note that both the flavour-diagonal and mixed-flavour contributions correspond to the $\gamma Z$ interference since there is no flavour change for the $uu$, $dd$ and $ud$ contributions. For the $ud$ piece, unlike the naming convention might suggest, the currents act on the individual $u$- and $d$-quarks in the nucleon. 
 
The asymptotic behaviour of the perturbed two-point correlator at large Euclidean time takes the familiar form,
\begin{equation} \label{eq:G2spec}
      G^{(2)}_{\bslam}(\pb,\qb,t) \simeq A_\bslam(\pb,\qb) e^{ - E_{N_{\bslam}}(\pb,\qb) \, t },
\end{equation}
where $E_{N_{\bslam}}$ is the perturbed energy of the ground state nucleon in the presence of an external field and $A_\bslam$ the corresponding overlap factor. In order to extract the second-order energy shift, we calculate the following combination of perturbed energies,
\begin{align} 
	\begin{split}
		\Delta E_{N_{\bslam}}(\pb,\qb) = \frac{1}{4} \Big[
		&E_{N_{(+\lambda_1, +\lambda_2)}}(\pb,\qb) +
		E_{N_{(-\lambda_1, -\lambda_2)}}(\pb,\qb) \\ 
		- &E_{N_{(+\lambda_1, -\lambda_2)}}(\pb,\qb) -
		E_{N_{(-\lambda_1, +\lambda_2)}}(\pb,\qb) \Big],
	\end{split}
\end{align} 
where this corresponds to the term of interest,
\begin{align} \label{eq:enshift_oo}
	\begin{split}
	    \Delta E_{N_{\bslam}}(\pb,\qb) &= \lambda_1 \lambda_2 \left. \frac{\partial^2 E_{N_{\lambda}}(\pb,\qb)}{\partial \lambda_1 \partial \lambda_2} \right|_{\bslam=0} \\
	    & + \mathcal{O}(\lambda_1 \lambda_2^3) + \mathcal{O}(\lambda_1^3 \lambda_2),
    \end{split}
\end{align}
appearing on the LHS of \Cref{eq:secondorder_fh} associated with the interference of the currents $\mathcal{J}_\mu$ and $\mathcal{J}_\nu^{A}$. In terms of perturbed correlators, the second-order energy shift is isolated via the ratio,
\begin{align} \label{eq:ratio}
	\begin{split}
		\mathcal{R}_{\bslam}(\pb,\qb,t) &\equiv 
		  \frac{
		  G^{(2)}_{+\lambda_1,+\lambda_2}(\pb,\qb,t) \, 
		  G^{(2)}_{-\lambda_1,-\lambda_2}(\pb,\qb,t)
		  }
		  {
		  G^{(2)}_{+\lambda_1,-\lambda_2}(\pb,\qb,t) \, 
		  G^{(2)}_{-\lambda_1,+\lambda_2}(\pb,\qb,t)
		  } \\
		  &\xrightarrow{t \gg 0} A^\prime_\bslam(\pb,\qb) e^{-4\Delta E_{N_\bslam}(\pb,\qb) \, t},
	\end{split}
\end{align} 
where we have collected all overlap factors into a single constant $A^\prime_\bslam(\pb,\qb)$ which is irrelevant for our discussion of energy shifts. This ratio isolates the energy shift only at even orders of $\bslam$, e.g. $\mathcal{O}(\lambda_1^n \lambda_2^m)$ with $n+m$ even and $n,m \ge 0$.

\section{Calculation and analysis details} \label{sec:simu}
Our calculations are performed on QCDSF's 2$+$1-flavour gauge configurations generated with a stout-smeared non-perturbatively $\mathcal{O}(a)$-improved Wilson action for the dynamical up/down and strange quarks and a tree-level Symanzik improved gauge action~\cite{Cundy:2009yy}. We use two ensembles with volume $V=48^3 \times 96$, and couplings $\beta=[5.65, \, 5.95]$ corresponding to the lattice spacings $a=[0.068, \, 0.052] \; {\rm fm}$, respectively. Quark masses are tuned to the $SU(3)$ symmetric point where the masses of all three quark flavours, $u$, $d$, and $s$, are set to approximately the physical flavour-singlet mass, $\overline{m} = (m_s + 2 m_l)/3$~\cite{Bietenholz:2010jr,Bietenholz:2011qq}, yielding $m_\pi \approx 415 \, {\rm MeV}$. Details of the gauge configurations we employ are listed in \Cref{tab:gauge_details}. We carry out our calculations with several values of current momenta, $Q^2$, in the range $0.5 \lesssim Q^2 \lesssim 10 \, {\rm GeV}^2$ (see \Cref{tab:qmom}). A valence quark propagator is computed for each combination of $\bslam$ and $\qb$. To increase statistics (mostly for low-$Q^2$ values), we perform up to $\mathcal{O}(10^3)$ measurements by employing up to two sources on the ensembles of size $\mathcal{O}(500)$ configurations.
\begin{table*}[t]
	\centering
	\caption{ \label{tab:gauge_details} Details of the gauge ensembles used in this work.}
	\setlength{\extrarowheight}{2pt}
	\begin{tabularx}{\textwidth}{C{1}C{1}C{1}C{1}C{1}C{1}C{1}C{1}C{1}C{1}C{1}C{1}}
		\hline\hline
		$N_f$ & $\beta$  & $a \, [{\rm fm}]$ & $c_{SW}$ & $\kappa_l$ & $\kappa_s$ & $L^3 \times T$ & $m_\pi \, [{\rm MeV}]$ & $m_\pi L$ & $Z_V$ & $Z_A$ & $N_{\rm cfg}$\\
		\hline
		$2+1$ & 5.65 & 0.068 & 2.48 & 0.122005 & 0.122005 & $48^3\times96$ & $412$ & $6.9$ & 0.8615 & 0.8754 & 535 \\
		$2+1$ & 5.95 & 0.052 & 2.22 & 0.123460 & 0.123460 & $48^3\times96$ & $418$ & $5.3$ & 0.8856 & 0.8983 & 507 \\
		\hline\hline
	\end{tabularx}
\end{table*}

To isolate the $\mathcal{F}_3$ term in \Cref{eq:f3}, we choose $\mu=1$, $\nu=3$, and $q_1=0$, $q_2 \ne 0$, for a nucleon at rest, $\pb=(0,0,0)$. This imposes the remaining indices to be $\alpha=4$ and $\beta=2$ to have a non-vanishing result, considering that $q_4=0$ in the Feynman-Hellmann approach. Hence we rewrite \Cref{eq:secondorder_fh} as
\begin{equation} \label{eq:f3ip}
    \frac{\mathcal{F}_3(\omega,Q^2)}{\omega} = \frac{Q^2}{q_2} \left. \frac{\partial^2 E_{N_{\bslam}}(\pb,\qb)}{\partial \lambda_1 \partial \lambda_2} \right|_{\bslam=0}.
\end{equation}  
Here we remind the reader that in practice, we determine the RHS of \Cref{eq:f3ip} which has no singularity at $\omega=0$. A further note is on the kinematic factor, which is obtained through a continuum formulation, however, we replace this factor with its lattice equivalent as will be discussed later on. 

We determine the energy shift, $\Delta E_{N_{\bslam}}(\pb,\qb)$, by performing a fit to the plateau region of the ratio given in \Cref{eq:ratio} for each $\qb$ excluding those indicated by italics in \Cref{tab:qmom}. Italicised $\qb$-momenta are analysed via direct fits to the perturbed correlators as described later. We note that only the connected diagrams are calculated for the correlators since the disconnected contribution vanishes for the leading-twist part of the lowest moment of $F_3$.
\begin{table}[t]
\centering
\caption{\label{tab:qmom} Multiple $\bv{q} = (q_1,q_2,q_3)$ values that we consider in this work given in lattice and physical units. Note that we have set $q_1 = 0$. The momenta for which we perform a multi-exponential analysis as described in \Cref{sec:fh} are indicated by italics.}
\setlength{\extrarowheight}{2pt}
\begin{tabularx}{.48\textwidth}{C{1}C{1}|C{1}C{1}}
	\hline\hline
	\multicolumn{2}{c|}{$\beta=5.65$} & \multicolumn{2}{c}{$\beta=5.95$} \\
	$\qb \, [2\pi/L]$ & $Q^2 \, [{\rm GeV}^2]$ & $\qb \, [2\pi/L]$ & $Q^2 \, [{\rm GeV}^2]$ \\
	\hline
	---       & ---  & {\em (0,1,1)} & 0.49 \\
	{\em (0, 2, 0)} & 0.57 & (0,2,0) & 0.98 \\
	(0, 1, 2) & 0.71 & (0,1,2) & 1.23 \\
	(0, 2, 1) & 0.71 & (0,2,1) & 1.23 \\
	(0, 2, 2) & 1.14 & (0,2,2) & 1.97 \\
	(0, 3, 0) & 1.29 & (0,1,3) & 2.46 \\
	(0, 1, 3) & 1.43 & (0,3,1) & 2.46 \\
	(0, 3, 1) & 1.43 & (0,2,3) & 3.20 \\
	(0, 2, 3) & 1.86 & (0,3,2) & 3.20 \\
	(0, 2, 4) & 2.86 & (0,1,4) & 4.18 \\
	(0, 3, 4) & 3.57 & (0,4,1) & 4.18 \\
	(0, 4, 3) & 3.57 & (0,3,3) & 4.43 \\
	(0, 5, 0) & 3.57 & (0,2,4) & 4.92 \\
	(0, 1, 5) & 3.71 & (0,4,2) & 4.92 \\
	(0, 3, 5) & 4.86 & (0,3,4) & 6.15 \\
	(0, 5, 3) & 4.86 & (0,4,3) & 6.15 \\
	(0, 2, 6) & 5.72 & (0,5,0) & 6.15 \\
	(0, 1, 7) & 7.14 & (0,2,5) & 7.13 \\
	(0, 7, 1) & 7.14 & (0,5,2) & 7.13 \\
	(0, 1, 8) & 9.29 & (0,1,6) & 9.10 \\
	(0, 4, 7) & 9.29 &   ---   & --- \\
	\hline\hline
\end{tabularx}
\end{table}
In order to control the uncertainty arising due to the choice of fit windows, $[t_{\rm min}, t_{\rm max}]$, we follow the weighted-averaging procedure outlined in~\cite{NPLQCD:2020ozd}. Several single exponential, \Cref{eq:G2spec}, fits are performed with fixed $t_{\rm max}$ and varying $t_{\rm min}$, where we demand $t_{\rm max}-t_{\rm min} \ge 3a$. The weight on each fit is calculated by~\cite{NPLQCD:2020ozd},
\begin{equation} \label{eq:weight}
 	w^f = \frac{p_f \, (\delta \mathcal{O}^f)^{-2}}{\sum_{f^\prime} p_{f^\prime} (\delta \mathcal{O}^{f^\prime})^{-2}},
\end{equation} 
where $p_f = 2 \operatorname{min}(CDF, 1-CDF)$ is the two-sided p-value of a $\chi^2$ distribution calculated with the cumulative distribution function $CDF(N_{\rm dof}/2, \chi^2_f/2) = \Gamma(N_{\rm dof}/2, \chi^2_f/2) / \Gamma(N_{\rm dof}/2)$. Here $f$ denotes the choice of fit window, and $\mathcal{O}$ is an unbiased estimator of a generic quantity, e.g. the energy shift extracted from the ratio. $\delta \mathcal{O}^f$ is the statistical uncertainty on $\mathcal{O}^f$ estimated by a bootstrap analysis. Lastly, the final estimate of the quantity of interest, $\bar{\mathcal{O}}$, and its uncertainty, $\delta \bar{\mathcal{O}}$ are calculated via,

\begin{subequations}
	\begin{align}
		\label{eq:wavg}
		\bar{\mathcal{O}} &= \sum_f w^f \mathcal{O}^f, \\ 
		\label{eq:wstat}
		\delta_{\rm stat} \bar{\mathcal{O}}^2 &= \sum_f w^f (\delta \bar{\mathcal{O}}^f)^2, \\
		\label{eq:wsys}
		\delta_{\rm sys} \bar{\mathcal{O}}^2 &= \sum_f w^f (\mathcal{O}^f - \bar{\mathcal{O}})^2, \\ 
		\label{eq:werr}
		\delta \bar{\mathcal{O}} &= \sqrt{\delta_{\rm stat} \bar{\mathcal{O}}^2 + \delta_{\rm sys} \bar{\mathcal{O}}^2}.
	\end{align}  
\end{subequations}

The quantity of interest in our case is the Compton amplitude, which is obtained from a fit to the $\lambda$ dependence of the energy shifts. To estimate the amplitude we calculate the ratio for $|\lambda_1| = |\lambda_2| = |\lambda|$, for two values of $|\lambda| = 0.0125$ and $0.025$, and perform fits of the form \Cref{eq:enshift_oo} to determine the second-order energy shift, $\left. \frac{\partial^2 E_{N_{\lambda}}(\pb)}{\partial \lambda_1 \partial \lambda_2} \right|_{\bslam=0}$, for each fit window, $f$, within their respective bootstrap samples. Evaluating two values of $|\lambda|$ allows us to assess the effect of higher-order terms in \Cref{eq:enshift_oo}. We provide a further stability check in \Cref{app:mexp}. Multiplied by the kinematic factor, \Cref{eq:f3ip}, the extracted second-order energy shift gives the Compton structure function. 

\begin{figure}[t]
    \centering
    \includegraphics[width=.48\textwidth]{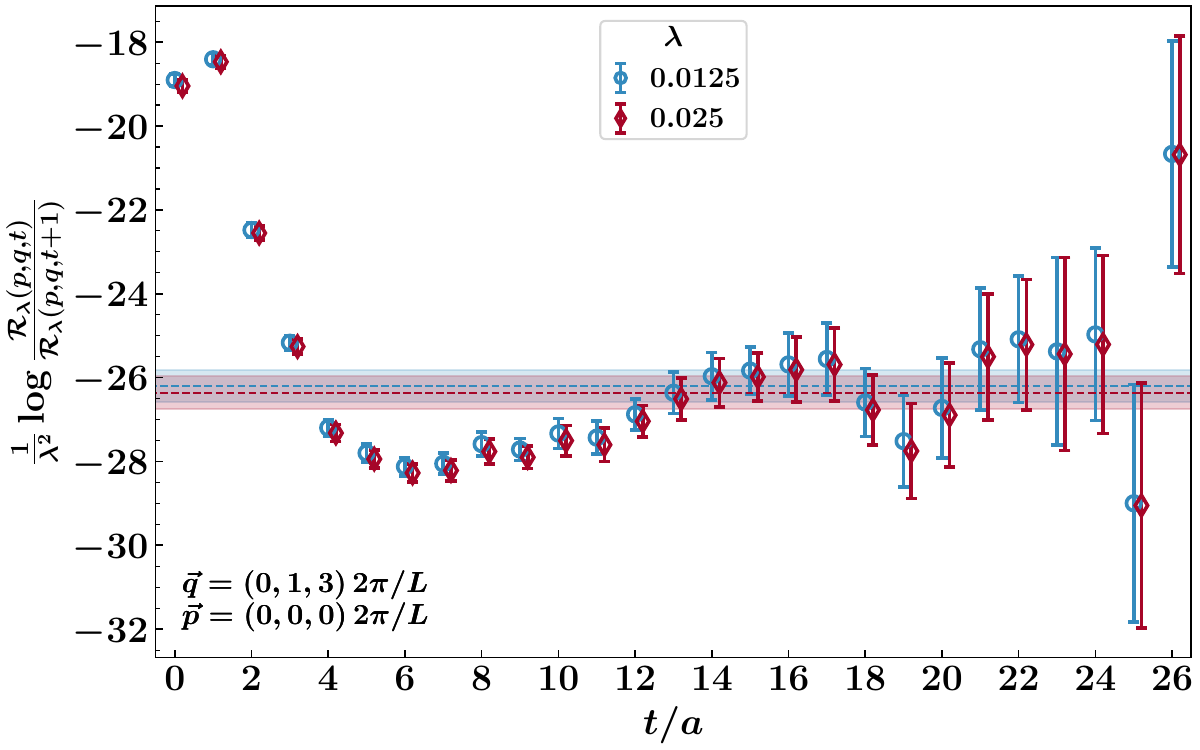}
    \caption{\label{fig:effmass}Effective energy plot analogue for the ratio, \Cref{eq:ratio}, normalised by $\lambda^2$ ($|\lambda_1|=|\lambda_2|=|\lambda|$, for $\lambda=0.0125$, and $0.025$). The extracted energy shifts, \Cref{eq:wavg}, and their uncertainties, \Cref{eq:werr}, are shown by the dashed lines and shaded bands respectively. We are showing the results for the $uu$ contribution obtained on the $\beta=5.95$ ensemble for $(\bv{p},\bv{q}) = ((0,0,0),(0,1,3)) \, \latmom$ corresponding to $\omega = 0$ at $Q^2 \sim 2.5 \, {\rm GeV}^2$. 
    }
\end{figure}

\begin{figure}[t]
    \centering
    \includegraphics[width=.45\textwidth]{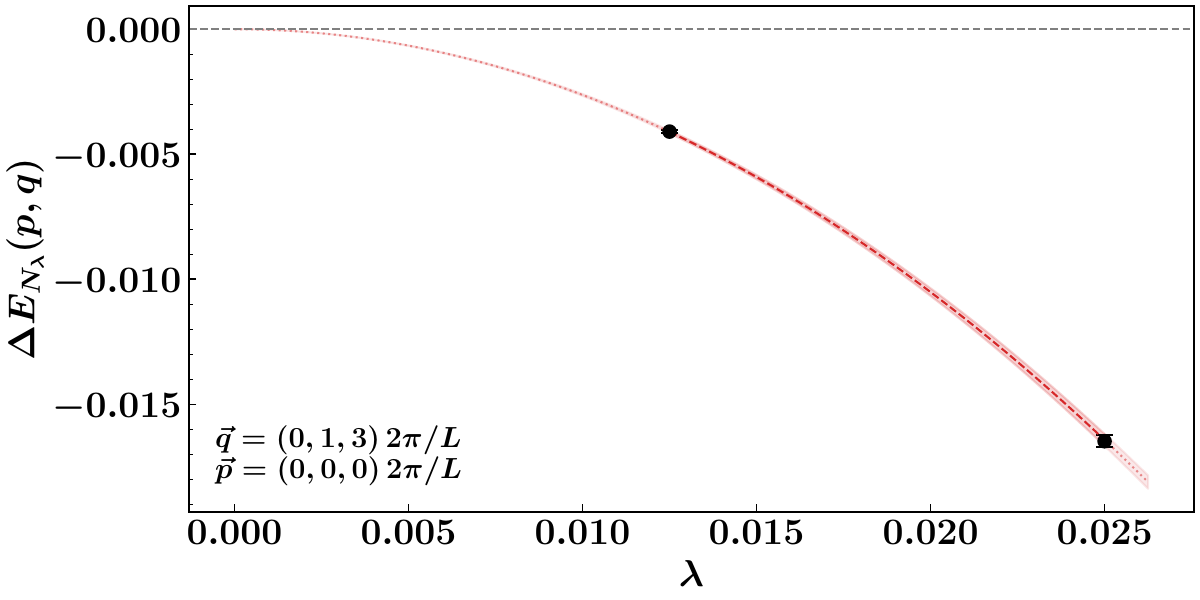}
    \caption{\label{fig:lamfit}$\lambda$ dependence of the energy shift for the same kinematics given in \Cref{fig:effmass}.
    }
\end{figure}

\begin{figure}[t]
    \centering
    \includegraphics[width=.48\textwidth]{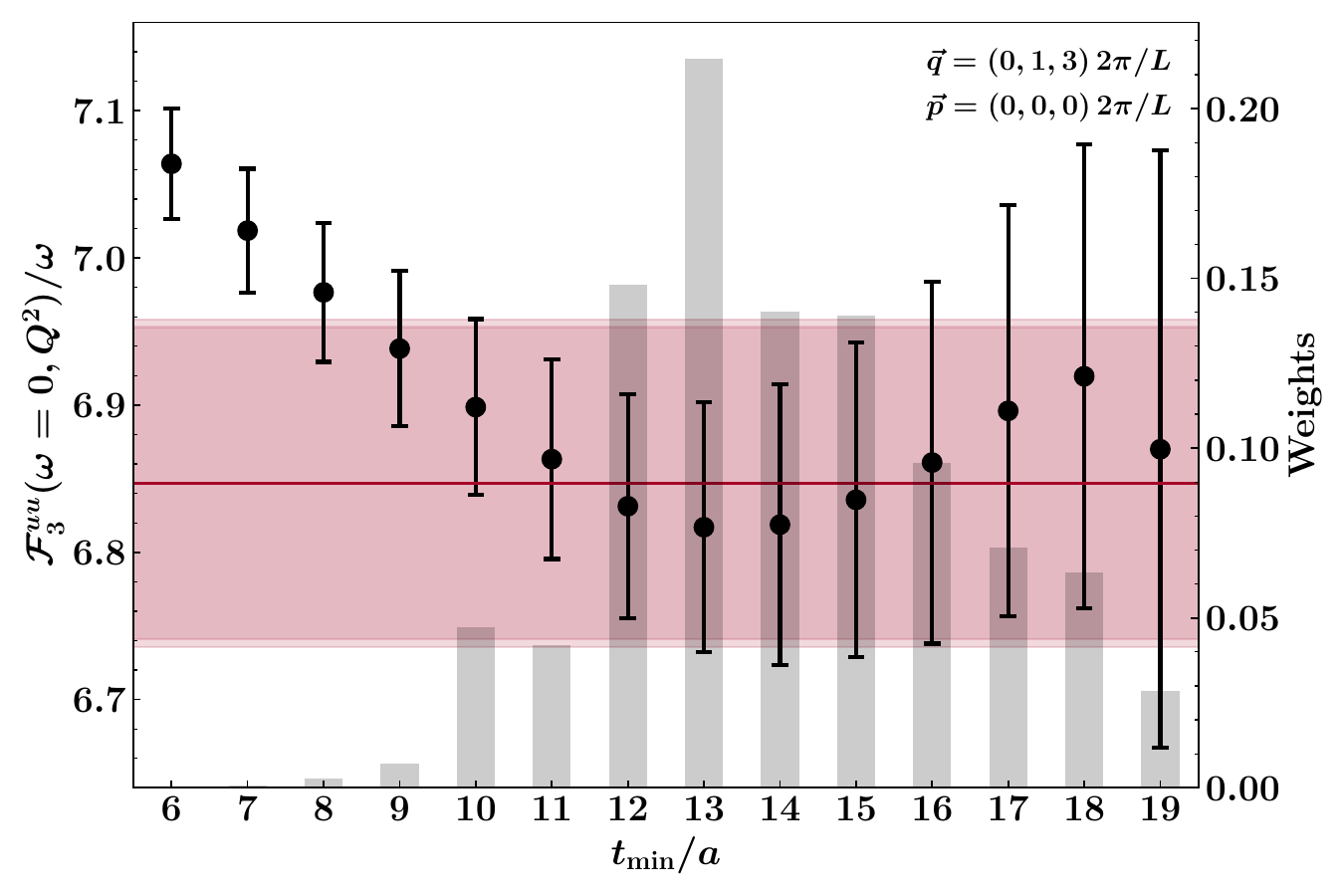}
    \caption{\label{fig:ooo_F3_uu_q013}The $\mathcal{F}_3$ Compton structure function obtained via \Cref{eq:f3ip} for the same kinematics given in \Cref{fig:effmass}. The $uu$ contribution determined on the $\beta=5.95$ ensemble is shown. $t_{\rm max}=24a$ for this case and we dropped the $t_{\rm min} > 19a$ points for clarity since their weights vanish due to low fit quality. The bars associated with the right y-axis denote the weight, \Cref{eq:weight}, of each point. The final estimate, \Cref{eq:wavg}, is shown by the solid line, and the inner and outer shaded bands denote the statistical, \Cref{eq:wstat}, which is dominant, and the total, \Cref{eq:werr}, uncertainties.
    }
\end{figure}

A representative effective energy plot analogue for the ratio for a nucleon at rest, $\pb = (0,0,0) \latmom$, at $\qb = (0,1,3) \latmom$ is shown in \Cref{fig:effmass}. The corresponding $\lambda$ fit is shown in \Cref{fig:lamfit}. 
We have plotted the weighted-average results in \Cref{fig:effmass,fig:lamfit} to illustrate the intermediate steps, but in practice we carry the analysis through the bootstrap samples until estimating the amplitude. In propagating the uncertainty for weighted averaging, we use the p-values calculated with the $\chi^2$ of the ratio fits for a fixed $\lambda$. Since we employ two $\lambda$ values in our analysis, choosing the p-values associated with any one of the $\lambda$ presents an additional source of systematic error. However, we do not find any significant difference among the p-values with respect to $\lambda$. On this note, we show the $\mathcal{F}_3$ Compton structure function as calculated using the energy shifts extracted on several fit windows, along with the weight of each estimate in \Cref{fig:ooo_F3_uu_q013}. 

While the ratio method described in \Cref{sec:fh} and further discussed in this section is the preferred method and utilised for the majority of the calculation, we follow a different approach for the $\qb$-momenta indicated by italics in \Cref{tab:qmom} for which the ratios do not exhibit a clear plateau. In those cases, we extract the perturbed energies directly from the perturbed correlators via a Bayesian approach using a multi-exponential fit ansatz,
\begin{equation} \label{eq:G2spec_3exp}
      G^{(2)}_{\bslam}(\pb,\qb,t) = \sum_{i=0}^2 A^{(i)}_\bslam(\pb,\qb) e^{ - E^{(i)}_{N_{\bslam}}(\pb,\qb) \, t },
\end{equation}
where $E^{(0)}_{N_{\bslam}}(\pb,\qb)$ is the perturbed ground state energy we are interested in. We choose flat priors for $A^{(0)}_\bslam$ and $E^{(0)}_{N_{\bslam}}$ with support in range $[0,1]$, and truncated Gaussian priors with support for positive real numbers only for $A^{(i)}_\bslam$ and $E^{(i)}_{N_{\bslam}}$ centred at zero with widths $\sigma=1$. A hierarchy of energies, $E^{(2)}_{N_{\bslam}} > E^{(1)}_{N_{\bslam}} > E^{(0)}_{N_{\bslam}} > 0$, is enforced by choosing the energy priors of consecutive states to be bounded from below by the previous energy state. No hierarchy is assumed for the overlap factors $A^{(i)}_\bslam$.

Once the individual perturbed ground state energies are extracted for the correlators appearing in \Cref{eq:ratio}, we determine the energy shift via,
\begin{align} \label{eq:mexp_enshift_oo}
	\begin{split}
		\Delta E_{N_{\bslam}}(\pb,\qb) &= \frac{1}{4} \Big[
	E^{(0)}_{N_{(+\lambda_1, +\lambda_2)}}(\pb,\qb) +
	E^{(0)}_{N_{(-\lambda_1, -\lambda_2)}}(\pb,\qb) \\ 
	&- E^{(0)}_{N_{(+\lambda_1, -\lambda_2)}}(\pb,\qb) -
	E^{(0)}_{N_{(-\lambda_1, +\lambda_2)}}(\pb,\qb) \Big],
	\end{split}
\end{align} 
where the unperturbed ground state energy, $E^{(0)}_N(\pb)$, is also extracted in the same Bayesian analysis framework. Further details of this approach and a comparison to the ratio method are collected in \Cref{app:mexp}.

\begin{figure}[t]
    \centering
    \includegraphics[width=.48\textwidth]{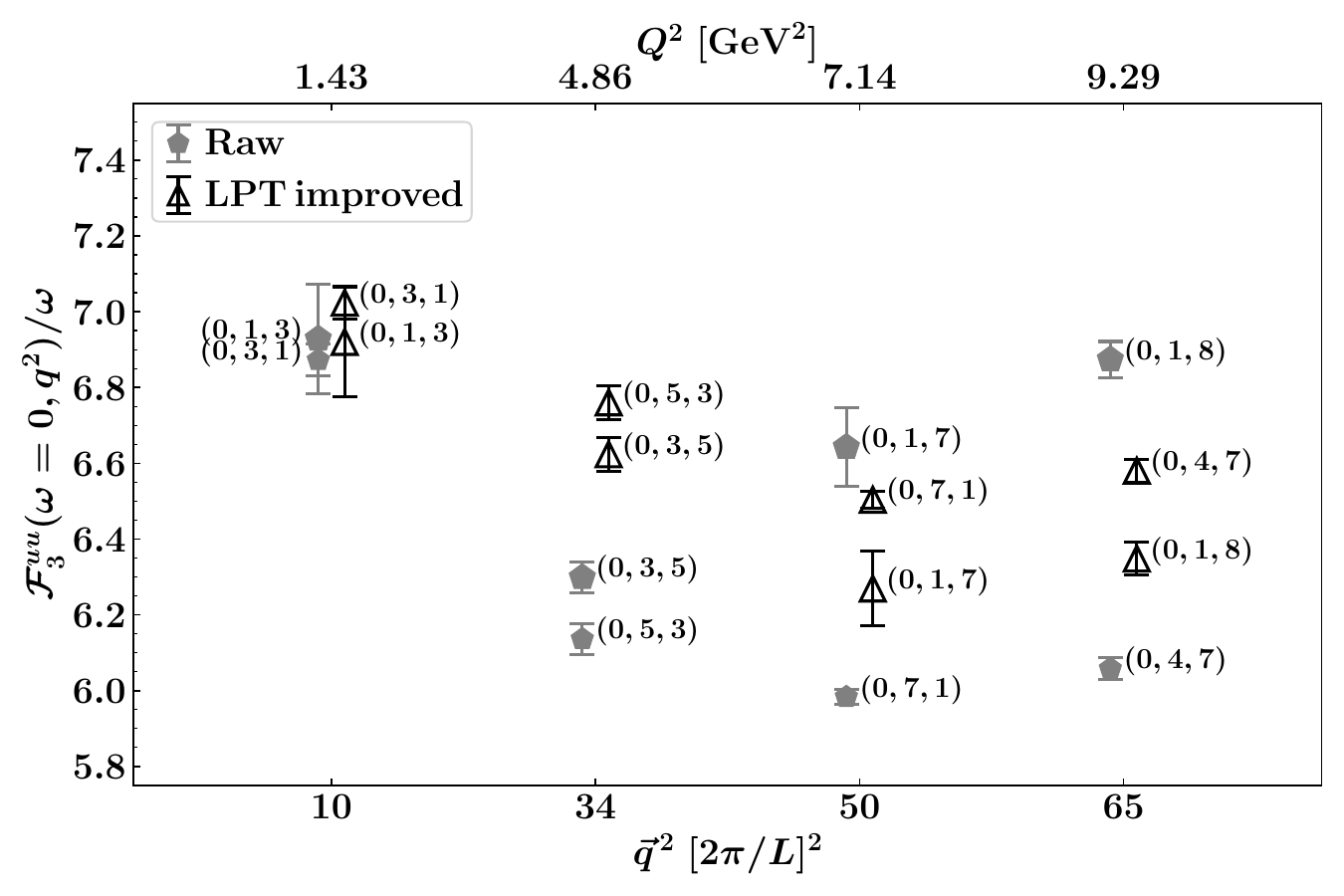}
    \caption{\label{fig:lpt_corr}The raw (filled pentagons) and LPT-improved (empty triangles) $\mathcal{F}_3$ Compton structure function at $\omega=0$ for a few representative kinematics obtained on the $\beta=5.65$ ensemble for the $uu$ contribution. We label the points with their corresponding $\qb = (q_1,q_2,q_3)$. Note that the $\qb^2$ is not evenly distributed, as indicated by the non-linear scale on the horizontal axis. Points are shifted for clarity.
    }
\end{figure}

Before we present our results, we remark on our reduction of discretisation errors arising due to broken rotational symmetry on the lattice. The Compton amplitude is related to the Compton structure function $\mathcal{F}_3$, and consequently to the moments of $F_3$, by the kinematic factor given in \Cref{eq:f3ip}. However, this relation starts from a continuum expression, \Cref{eq:f3}, and naturally receives corrections in a lattice formulation. Based on a lattice perturbation theory (LPT) inspired procedure~\cite{Gockeler:2006nq,Tom:2024tgs}, we derive a correction factor for a free fermion (see \Cref{app:lpt}) and apply this correction to obtain the LPT-improved structure function at $\omega = 0$. The improvement is achieved by replacing the kinematic factor,
\begin{equation}\label{eq:lpt_corr}
	\frac{Q^2}{q_2} \to \frac{\sum_i \sin^2 q_i + \left[ \sum_i (1 - \cos q_i)\right]^2}{\sin q_2},
\end{equation}
appearing in \Cref{eq:f3ip}. We show the effect of this improvement in \Cref{fig:lpt_corr} for a few illustrative cases. At fixed $q^2$, the Compton structure function calculated using different $q$ geometries will receive different discretisation errors and hence will give different results as apparent for the raw points in \Cref{fig:lpt_corr}, and especially pronounced for the larger $q^2$. In the continuum limit however, using different $q$ geometries should lead to the same result. The LPT improved points in \Cref{fig:lpt_corr} show the correction factor works to reduce the effect of this discretisation error due to broken rotational symmetry, however residual discretisation artefacts are still evident.     

\section{Results and discussion} \label{sec:res}
Our main result is the lowest odd Mellin moment of $F_3$ determined following the analysis procedure outlined in \Cref{sec:simu} from the $\mathcal{F}_3$ Compton structure function at $\omega=0$,
\begin{equation}
	M_{1,qq}^{(3)}(Q^2) = \lim_{\omega \to 0} \frac{\mathcal{F}_3^{qq}(\omega=0,Q^2)}{4 \, \omega},
\end{equation}
for each $Q^2$ given in \Cref{tab:qmom}. Here, $qq$ labels the flavour-diagonal $uu$ or $dd$ pieces. 

\begin{figure*}[t]
    \centering
    \includegraphics[width=0.7\textwidth]{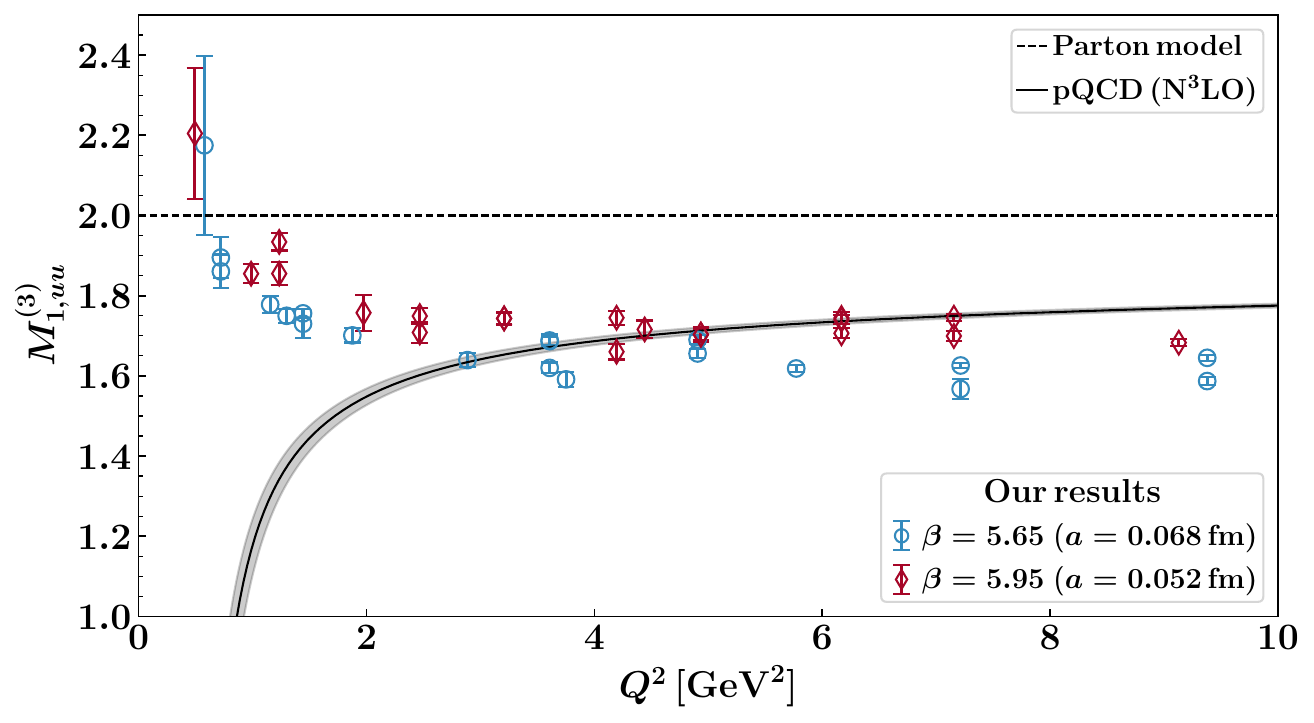} \\
    \includegraphics[width=0.7\textwidth]{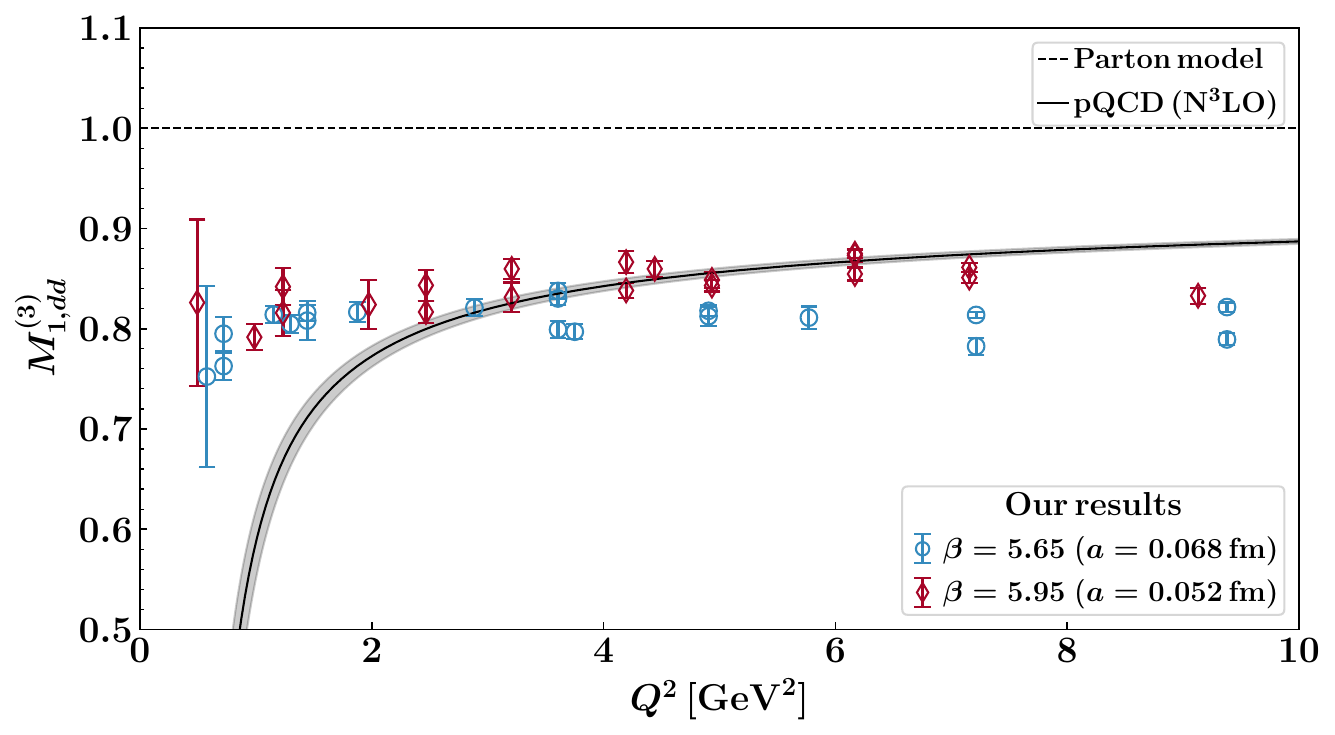}
    \caption{\label{fig:m1_F3}The $uu$ and $dd$ contributions to the lowest odd Mellin moment of the $F_3$ structure function as a function of $Q^2$. We compare our lattice results from coarse (blue circles) and fine ensembles (red diamonds) to the perturbative QCD (pQCD) predictions estimated to N${}^3$LO (\Cref{eq:GLS_n3lo} without the higher-twist term).
    }
\end{figure*}

We show our LPT-improved moments obtained on the two lattice ensembles in \Cref{fig:m1_F3} for the $uu$ and $dd$ contributions. For the first time we see lattice moments for the GLS sum rule covering a wide range of $Q^2$ from the perturbative to the nonperturbative regime. Therefore it is instructive to consider different $Q^2$ regions where different physical effects are at play, although the boundaries between these regions are not definite. 

The high-$Q^2$ region, $Q^2 \gtrsim 4 \, {\rm GeV}^2$, is dominated by the leading-twist parton distribution functions and well-described by perturbative QCD. In the quark-parton model, the lowest odd moment of $F_3$ is given by ~\cite{ParticleDataGroup:2024cfk},
\begin{align}\label{eq:gZ_QPM}
	\begin{split}
		M_{1,qq}^{(3), {\rm QPM}}(Q^2) &= \int_0^1 dx \, (q-\bar q)(x,Q^2) \\ 
		&= N_q,
	\end{split}
\end{align} 
where $q(x,Q^2)$ and $\bar q(x,Q^2)$ denote the leading-twist quark and anti-quark parton distribution functions, respectively, and therefore $N_q$ is the number of valence quarks of flavour $q$.

In QCD, the quark-parton model value receives perturbative corrections which is calculated to N${}^3$LO~\cite{Larin:1991tj,Hayen:2020cxh},
\begin{equation} \label{eq:GLS_n3lo}
	M_{1,qq}^{(3)}(Q^2) = N_q \left[ 1 - \sum_{i=1}^4 C_i \, a_s^i(Q^2) \right] + \frac{\Delta^{\rm HT}_q}{Q^2} + \cdots,
\end{equation}
where $N_q = 2$ and $1$ for the $uu$ and $dd$ contributions respectively, the coefficients are
$C_1 = 1$, 
$C_2 = \frac{55}{12}-\frac{1}{3} n_f$, 
$C_3 = 41.440 - 8.020 n_f + 0.177 n_f^2$, and
$C_4 = 479.4 - 117.6 n_f + 7.464 n_f^2 - 0.1037 n_f^3$~\cite{Larin:1991tj,Hayen:2020cxh}, where we take $n_f=3$,
$a_s(Q^2) \equiv \alpha_s(Q^2)/\pi$, and $\Delta^{\rm HT}$ denote the coefficient of the leading higher-twist term with ellipses denoting the neglected higher-order power corrections. Note that there are additional corrections to the GLS sum rule arising from strange quark distributions and from charge symmetry violating parton distributions~\cite{Londergan:2010cd}, and heavy flavours~\cite{Blumlein:2016xcy}, however, we do not incorporate these corrections to our current analysis at this stage. The $\gamma Z$ analogue of the GLS sum rule for $\nu$DIS can be constructed via,
\begin{equation}
	M_{1,qq}^{(3),\gamma Z}(Q^2) = \sum_{q=u,d} 2 \mathcal{Q}_q g_A^q M_{1,qq}^{(3), {\rm QPM}}(Q^2),
\end{equation}
where $\mathcal{Q}_q$ and $g_A^q$ are the electric and axial charges of the quark $q$. 

The scale dependence of this expression is trivial, i.e. it is carried by the scaling of $\alpha_s$ only, given that the anomalous dimensions of the $\gamma_\mu$ operator vanishes at leading order in operator product expansion~\cite{Retey:2000nq}. The solid bands in \Cref{fig:m1_F3} show the leading twist moment with perturbative corrections, \Cref{eq:GLS_n3lo}, calculated using $\alpha_s(Q^2)$ estimated to 4-loop accuracy in the ${\rm \overline{MS}}$ scheme following the PDG definition~\cite{ParticleDataGroup:2024cfk} and with the 3-flavour $\Lambda_{\rm \overline{MS}}^{(3)} = 338(10) \; {\rm MeV}$ quoted by the Flavour Lattice Averaging Group~\cite{FLAG:2024oxs} based on the determinations from Refs. \cite{Maltman:2008bx,PACS-CS:2009zxm,McNeile:2010ji,Chakraborty:2014aca,Bruno:2017gxd,Bazavov:2019qoo,Ayala:2020odx,Cali:2020hrj,DallaBrida:2022eua,Petreczky:2020tky}. 

Our moments are in good agreement with the perturbative QCD curve in the perturbative region. The leading-twist matrix element in the operator product expansion is of the $\gamma_0$ operator, which is reflected in the first moment being proportional to the number of valence quarks. Therefore we do not expect to have a quark-mass dependence in our lattice moments since $N_q$ does not depend on the quark masses. Then, in principle, these moments can be used to estimate the strong coupling constant at a fixed $Q^2$ by solving \Cref{eq:GLS_n3lo} for $\alpha_s$. Such an approach, however, would be premature before a continuum extrapolation since it does not provide a handle to control the mild residual discretisation artefacts that we observe. 

The mid-$Q^2$ region, $0.5 \lesssim Q^2 \lesssim 4 \, {\rm GeV}^2$, is relevant for higher-twist effects where they start to have a non-negligible contribution. The higher-twist corrections have been addressed in model calculations before~\cite{Braun:1986ty,Ross:1993gb,Dasgupta1996,Balla:1997hf}, and indicate $\Delta^{\rm HT} < 0$ for the GLS sum rule, albeit with low precision. We already see a clear deviation between our moments and the perturbative curve starting around $Q^2 \sim 4 \, {\rm GeV}^2$. The discrepancy between our results and the perturbative estimate indicate that we have $\Delta^{\rm HT}_q > 0$ in contrast to the model calculations.

\begin{figure*}[t]
    \centering
    \includegraphics[width=.8\textwidth]{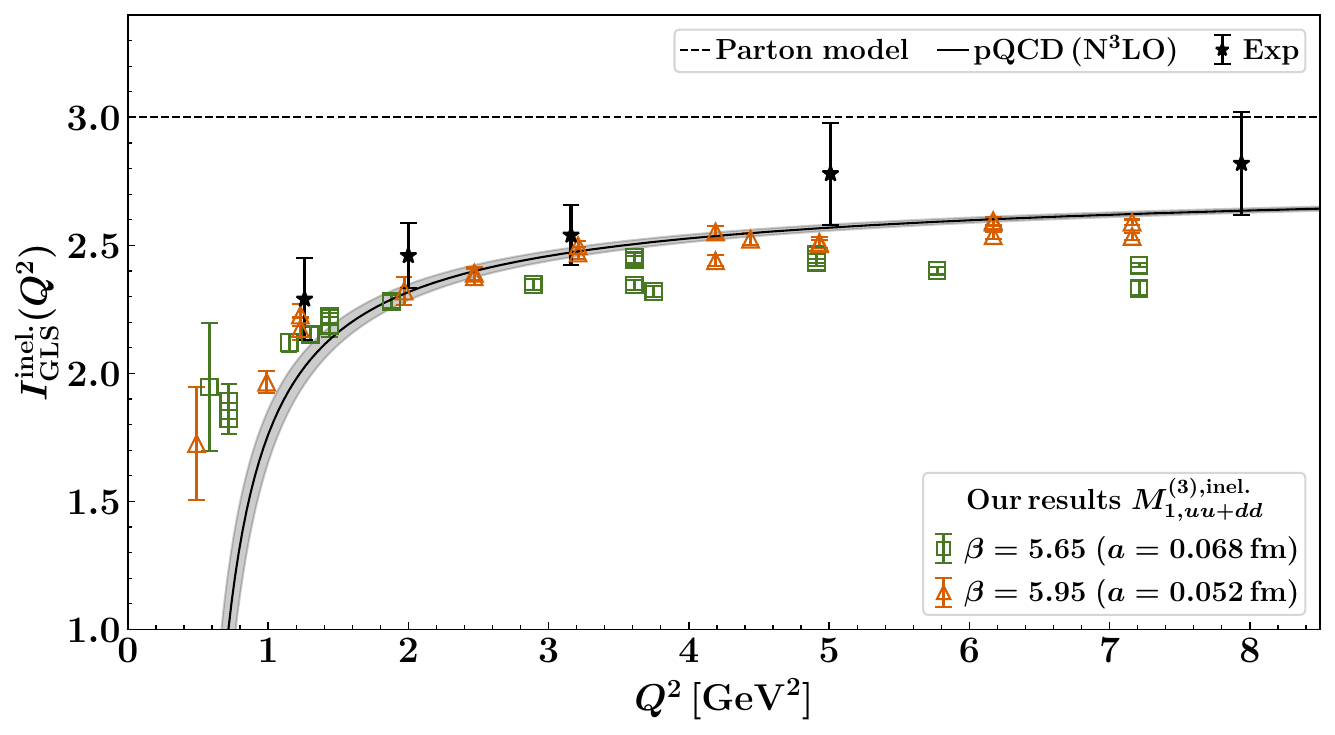} 
    \caption{\label{fig:m1_upd_F3}The inelastic part of the GLS sum rule, \Cref{eq:GLS}, as a function of $Q^2$. We compare our results, $M_{1,uu+dd}^{(3), {\rm inel.}}$, to the experimentally determined elastic-subtracted GLS sum rule~\cite{Kim:1998kia}, and the parton model and perturbative QCD (pQCD) predictions estimated to N${}^3$LO (\Cref{eq:GLS_n3lo} with $N_q=N_u+N_d$ and higher-twist terms neglected).
    }
\end{figure*}

Lower $Q^2$ values are expected to be dominated by the elastic pole~\cite{Blunden:2011rd},
\begin{align}\label{eq:f3_el}
	F_{3,q}^{\rm el.}(x,Q^2) = -G_A^{q}(Q^2) G_M^{q}(Q^2) \delta(1-x),
\end{align}
where $G_A^q$ and $G_M^q$ are the axial-vector and magnetic Sachs form factors, respectively, of the quark $q$. We adopt the normalisation $G_A^q(0) \equiv -g_A^q$ following PDG definitions~\cite{ParticleDataGroup:2024cfk}, where $g_A^q$ is the axial charge with $g_A^u > 0$ and $g_A^d < 0$. The $\delta(1-x)$ factor ensures that the only contribution to the dispersion integral, \Cref{eq:disp}, comes from $x=1$, hence for a nucleon at rest (the $\omega=0$ case), \Cref{eq:f3_el} gives the elastic contribution to the lowest moment, $M_{1}^{(3), \, {\rm el.}}(Q^2)$. 

Assuming a dipole parametrisation for both the axial-vector and magnetic form factors, the elastic contribution scales as $1/Q^8$, making it highly suppressed relative to the leading higher-twist contribution, however, it would still constitute a contamination. In order to estimate the leading $1/Q^2$ higher-twist contribution, the elastic contribution needs to be subtracted. We subtract the elastic contribution from our moments,
\begin{equation}
  	M_{1,qq}^{(3), \, {\rm inel.}}(Q^2) = M_{1,qq}^{(3)}(Q^2) - M_{1,qq}^{(3), \, {\rm el.}}(Q^2),
\end{equation}  
for $qq=uu$, $dd$, using the form factors determined on the same set of lattice ensembles~\cite{Batelaan:2022fdq,Smail:2023qlx} considering only the connected diagrams, since there are no contributions from disconnected diagrams to the leading-twist part of the $F_3$ structure function.. We adopt a dipole form for the axial-vector form factors and parametrise the Sachs magnetic form factors using a Pad\'e approximant following Kelly~\cite{Kelly:2004hm}. Further details of our calculations for the elastic contribution are given in \Cref{app:f3_el}. 

In order to make a comparison to experimental determinations of the GLS sum rule, we consider the $uu+dd$ moment, $M_{1,uu+dd}^{(3)}(Q^2) \equiv M_{1,uu}^{(3)}(Q^2) + M_{1,dd}^{(3)}(Q^2)$ and show our elastic-subtracted $uu+dd$ moments, $M_{1,uu+dd}^{(3), {\rm inel.}}$, in~\Cref{fig:m1_upd_F3}. The experimental measurements are taken from Ref.~\cite{Kim:1998kia} with the quasi-elastic contributions subtracted and quoted errors added in quadrature. Our results are in good agreement with the experimental points where the agreement gets better as the lattice spacing gets smaller. 

We note that the discrepancy between our elastic-subtracted moments and the perturbative curve towards the low-$Q^2$ region is a genuine higher-twist effect, modulo target-mass corrections. Our approach here paves the way for an extraction of the higher-twist coefficient from a fully nonperturbative calculation for each flavour. A caveat is that the higher-twist term in \Cref{eq:GLS_n3lo} is intricately connected to the divergence of the perturbative series, and great care needs to be taken in defining $\Delta^{\rm HT}$ for a rigorous estimation due to renormalon ambiguity mimicking a $\Lambda^2_{\rm QCD}/Q^2$ behaviour~\cite{Mueller:1993pa,Martinelli:1996pk,Beneke:1998ui,Beneke2000,Braun:2022gzl,Kronfeld:2023jab}. Qualitatively, our results indicate a positive higher-twist correction as opposed to the negative estimations of model calculations~\cite{Braun:1986ty,Ross:1993gb,Dasgupta1996,Balla:1997hf}.

\section{Summary} \label{sec:sum}
We provided a lattice QCD computation of the Gross-Llewellyn Smith sum rule by extracting the first moment of the parity-violating structure function of the nucleon from a direct calculation of the forward Compton amplitude based on a Feynman-Hellmann approach. Calculations are performed on two lattice ensembles with volume $48^3 \times 96$ and lattice spacings $a=0.068$ and $0.052$ fm. Both ensembles have the quark masses tuned to the $SU(3)$ symmetric point yielding $m_\pi \approx 415 \, {\rm MeV}$. Our results cover the $0.5 \lesssim Q^2 \lesssim 10 \; {\rm GeV}^2$ range, which include the nonperturbative and perturbative regions. In computing the Compton amplitude, we have considered a weighted averaging method to control the systematic errors arising due to the choice of fit windows in a standard two-point function analysis. Additionally, we have improved our results by correcting for one of the discretisation errors using a lattice perturbation theory estimation. We achieve percent-level statistical precision for our $Q^2 \gtrsim 1 \; {\rm GeV}^2$moments. Our results are compared to the Gross-Llewellyn Smith sum rule and a good agreement is obtained in the perturbative region, whereas we see clear deviations towards lower $Q^2$ values, indicating the presence of power corrections. We also find very good agreement with an experimental determination of the GLS sum rule. This work shows the feasibility of calculating the GLS sum rule on the lattice. A natural direction to take is quantifying the lattice systematics with further calculations using additional ensembles with varying lattice spacings, volumes and pion masses. Therefore, our calculation here provides an intermediate step towards i) studying the higher-twist effects in the GLS sum rule in more depth, ii) a determination of $\alpha_s$ from an hadronic observable, and iii) a direct ab-initio estimation of the electroweak box diagrams relevant for the determinations of $V_{ud}$ and the weak mixing angle, with controlled lattice systematic errors.

\acknowledgments
The numerical configuration generation (using the BQCD lattice QCD program~\cite{Haar:2017ubh})) and data analysis (using the Chroma software library~\cite{Edwards:2004sx}) was carried out on the Extreme Scaling Service (DiRAC, EPCC, Edinburgh, UK), the Data Intensive Service (DiRAC, CSD3, Cambridge, UK), the Gauss Centre for Supercomputing (NIC, Jülich, Germany), the NHR Alliance (Germany), and resources provided by the NCI National Facility in Canberra, Australia (supported by the Australian Commonwealth Government), the Pawsey Supercomputing Centre (supported by the Australian Commonwealth Government and the Government of Western Australia), and the Phoenix HPC service (University of Adelaide). RH is supported by STFC through grant ST/X000494/1. PELR is supported in part by the STFC under contract ST/G00062X/1. GS is supported by DFG Grant SCHI 179/8-1. JAC and TGS are supported by an Australian Government Research Training Program (RTP) Scholarship. KUC, RDY and JMZ are supported by the Australian Research Council grants DP190100297, DP220103098, and DP240102839. For the purpose of open access, the authors have applied a Creative Commons Attribution (CC BY) licence to any Author Accepted Manuscript version arising from this submission.

\appendix 
\section{Further details of the multi-exponential analysis and a comparison to the ratio method} \label{app:mexp}
As described in \Cref{sec:simu} we perform a multi-exponential analysis for the perturbed correlators to extract the second-order energy shift as opposed to the ratio method for two $Q^2$ points. In this section we provide further details for $\qb=(0,1,2) \latmom$ performed on the $\beta=5.95$ ensemble. The reason we choose this momentum is because the ratio method still gives reliable estimates and at the same time it is close to the momenta where the ratio method breaks down. Therefore, while we illustrate the multi-exponential analysis, we also provide a comparison of the ratio and multi-exponential methods to show that both give consistent results. Additionally, a further stability check is performed for the $\lambda$ fits by including two more $\lambda$ values to the analysis. 

\begin{figure}[h]
	\centering
	\includegraphics[width=.48\textwidth]{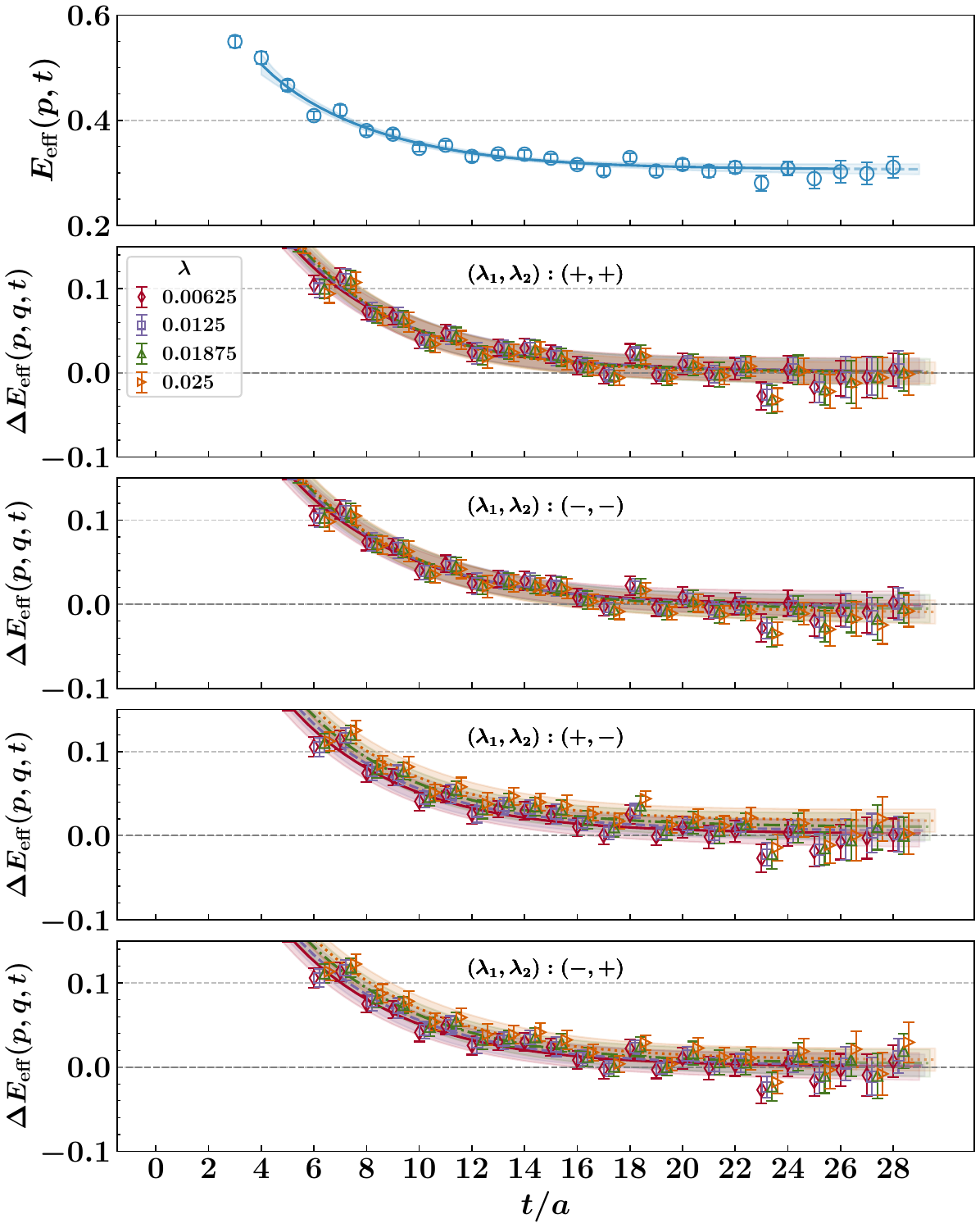}
	\caption{\label{fig:mexp_fits}Effective energy plot analogues for the multi-exponential fits to the unperturbed (top panel) and perturbed (bottom four panels) correlators. See the discussion in text for the perturbed results. The blue empty circles indicate the unperturbed correlator while the results for different $\lambda$ values are denoted by empty red diamonds ($\lambda=0.00625$), purple squares ($\lambda=0.0125$), green upward-oriented triangles ($\lambda=0.0375$), and orange right-oriented triangles ($\lambda=0.025$). }
\end{figure}

Using \Cref{eq:G2spec_3exp}, we determine the unperturbed and perturbed energies for $\lambda=[0.00625, 0.0125, 0.01875, 0.025]$ in a Bayesian analysis as described in \Cref{sec:simu}. Effective energy plot analogues for the resulting fits are shown in \Cref{fig:mexp_fits}. We show the effective energy for the unperturbed correlator in the top panel while the following panels show the effective energy difference obtained via,
\begin{equation}
  	\Delta G^{(2)}_{(\lambda_1, \lambda_2)}(\pb,\qb,t) = \frac{G^{(2)}_{(\lambda_1, \lambda_2)}(\pb,\qb,t)}{A^{(0)}(\pb) e^{ - E^{(0)}_{N}(\pb) \, t }}
\end{equation}  
for each perturbed correlator, $G^{(2)}_{(\lambda_1, \lambda_2)}$, that appear in \Cref{eq:ratio}. Here, $A^{(0)}$ and $E_N^{(0)}$ are the overlap factor and the energy of the unperturbed ground state nucleon determined from a multi-exponential fit to the unperturbed correlator respectively. We note that we fit \Cref{eq:G2spec_3exp} directly to the perturbed correlators and the effective energy differences shown in \Cref{fig:mexp_fits} are for a clearer illustration. The effective energy and effective energy difference are defined as,
\begin{align}
	E_{\rm eff}(\pb,t) &= \frac{1}{a} \log \frac{G^{(2)}(\pb,t)}{G^{(2)}(\pb,t+1)}, \\
	\Delta E_{\rm eff}(\pb,\qb,t) &= \frac{1}{a} \log \frac{\Delta G^{(2)}_{(\lambda_1, \lambda_2)}(\pb,\qb,t)}{\Delta G^{(2)}_{(\lambda_1, \lambda_2)}(\pb,\qb,t+1)},
\end{align}
respectively, where $G^{(2)}(\pb,t)$ is the unperturbed correlator.

Having determined the perturbed ($E^{(0)}_{N_{\lambda_1, \lambda_2}}(\pb, \qb)$) and unperturbed ($E^{(0)}_N(\pb)$) ground state energies, we calculate the energy shifts following \Cref{eq:mexp_enshift_oo} for each $\lambda$. Subsequently, we perform fits quadratic ($b \lambda^2$) and quadratic-plus-quartic ($b^\prime \lambda^2 + c \, \lambda^4$) in $\lambda$ to extract the second-order energy shift, $\left. \frac{\partial^2 E_{N_{\lambda}}(\pb)}{\partial \lambda_1 \partial \lambda_2} \right|_{\bslam=0}$, which is denoted by $b$ and $b^\prime$, respectively. 

\begin{table}[t]
	\centering
	\caption{ \label{tab:mexp_fitcomp} Details of the methodologies used to determine $M_{1,uu}^{(3)}$. For example, Fit 3 indicates that we have used four energy shifts, extracted via the ratio method for each $\lambda$ given in brackets, to perform the $\lambda$ fit using the quadratic-plus-quartic function, $b^\prime \lambda^2 + c \, \lambda^4$.}
	\setlength{\extrarowheight}{2pt}
	\begin{tabularx}{.48\textwidth}{L{.2} L{2.3} C{0.7} C{.7}}
		\hline\hline
		Fit & $\bslam$ & function  & method \\
		\hline
		1. & [0.0125, 0.025] & $b \, \lambda^2$ & ratio \\
		2. & [0.00625, 0.0125, 0.01875, 0.025] & $b \, \lambda^2$ & ratio \\
		3. & [0.00625, 0.0125, 0.01875, 0.025] & $b^\prime \lambda^2 + c \, \lambda^4$ & ratio \\
		4. & [0.0125, 0.025] & $b \, \lambda^2$ & multi-exp \\
		5. & [0.00625, 0.0125, 0.01875, 0.025] & $b \, \lambda^2$ & multi-exp \\
		6. & [0.00625, 0.0125, 0.01875, 0.025] & $b^\prime \lambda^2 + c \, \lambda^4$ & multi-exp \\
		\hline\hline
	\end{tabularx}
\end{table}

\begin{figure}[h]
	\centering
	\includegraphics[width=.48\textwidth]{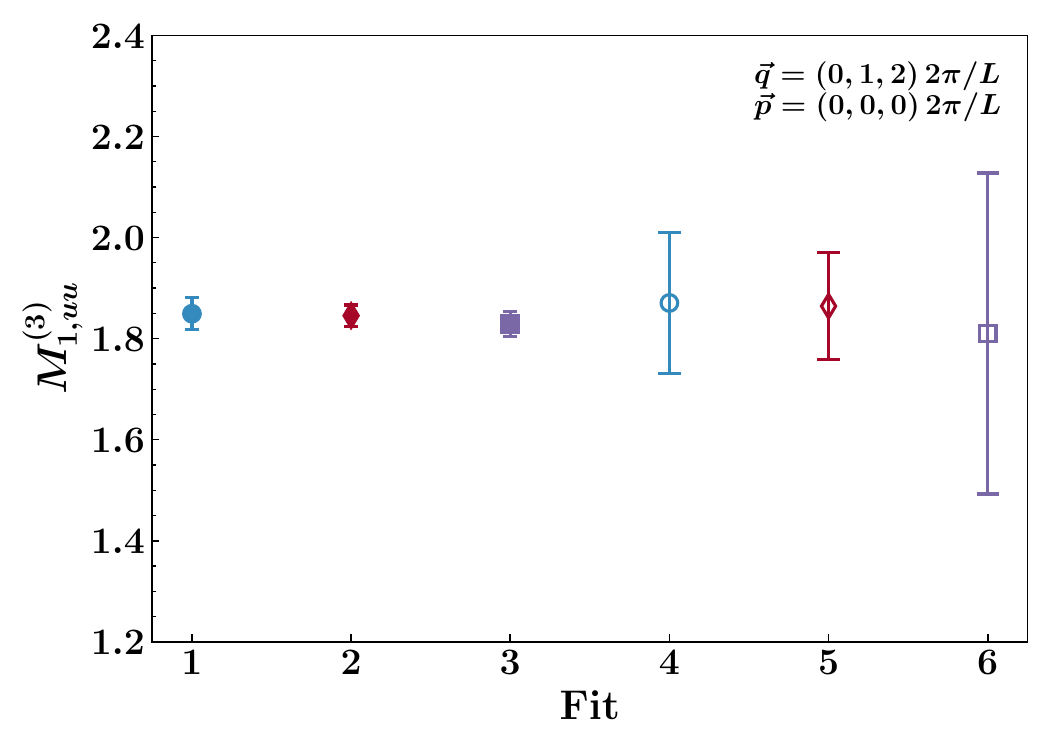}
	\caption{\label{fig:mexp_fitcomp}The $M_{1,uu}^{(3)}$ moment determined using methodologies listed in \Cref{tab:mexp_fitcomp}. We show the results for $\qb=(0,1,2) \latmom$ obtained on the $\beta=5.95$ ensemble.}
\end{figure}

In \Cref{fig:mexp_fitcomp} we show the $M_{1,uu}^{(3)}$ moment determined from the second-order energy shifts calculated via the methodologies listed in \Cref{tab:mexp_fitcomp}. Fits $1-3$ utilise the ratio method discussed in \Cref{sec:fh} to extract the energy shifts from a single fit window fixed for all $\lambda$ values. We see a good agreement between all determinations. The agreement between the Fits 1 and 3 (or 2 and 3) is within $1\sigma$ uncertainty which shows there is no statistically significant contamination due to the neglected $\mathcal{O}(\lambda_1 \lambda_2^3)$ and $\mathcal{O}(\lambda_1^3 \lambda_2)$ terms in \Cref{eq:enshift_oo}. The same statement holds for the multi-exponential fits as well. 

We conclude that the multi-exponential analysis gives consistent results in comparison to the ratio method and can be used to reliably determine the energy shifts when the ratio method fails albeit with increased statistical uncertainty.

\section{Details of lattice perturbation theory calculation} \label{app:lpt}
\begin{figure}[h]
	\centering
	\includegraphics[width=.48\textwidth]{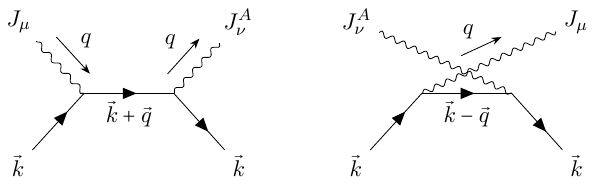}
	\caption{\label{fig:T_lpt}Diagrams relevant for Compton scattering.}
\end{figure}

We follow the work presented in~\cite{Gockeler:2006nq,Tom:2024tgs} to derive the correction factor discussed in the main text. To make a connection to the Lorentz decomposition of the Compton tensor (\Cref{eq:f3}) we work in Minkowski space in this section. On the lattice, using Wilson fermions with $r=1$, the tree-level Compton amplitude of a quark for mixed local vector and axial-vector currents $J_\mu = \bar{q} \gamma_\mu q$ and $J_\mu^A = \bar{q} \gamma_5 \gamma_\mu q$, respectively, is given by,
\begin{align} \label{eq:T_lpt}
	\begin{split}
		T_{\mu\nu}^{VA}(k,q) &= \bar{u}(k,s^\prime) \gamma_5 \gamma_\nu D_W^{-1}(k,q) \gamma_\mu u(k,s) \\ &+ (\mu \leftrightarrow \nu , q \to -q),
	\end{split}
\end{align}
where we have defined the propagator as the inverse of the Dirac term,
\begin{align}
	&D_W(k,q) = i \gamma_\rho \sin{(k+q)_\rho} + \bar{m}(k,q), \\
	&\bar{m}(k,q) = \sum_\rho [1 - \cos{(k+q)_\rho}] + m,
\end{align}{}
with $m$ the mass of the quark. The terms in \Cref{eq:T_lpt} correspond to the normal and crossed diagrams shown in \Cref{fig:T_lpt} respectively. 

Expanding for small $m$, and making use of the formula for the inverse of linear combinations of gamma matrices,
\begin{equation}
	(i \gamma_\mu b_\mu + a \mathbb{I})^{-1} = \frac{-i \gamma_\mu b_\mu + a \mathbb{I}}{\sum_\mu b_\mu^2 + a \mathbb{I}},
\end{equation}
we rewrite the propagator,
\begin{equation}
	D_W^{-1}(k,q) = \frac{\bar{D}_W(k,q)}{\tilde{Q}^2(k,q)} + \mathcal{O}(m),
\end{equation}
where,
\begin{align}
	& \bar{D}_W(k,q) = i \gamma_\rho \sin{(k+q)_\rho} + \sum_\rho [1 - \cos{(k+q)_\rho}], \\
	& \tilde{Q}^2(k,q) = \sum_\rho \sin^2{(k+q)_\rho} + \left( \sum_\rho [1 - \cos{(k+q)_\rho}] \right)^2.
\end{align}

We can simplify the gamma matrix combinations in the numerator using the relations, 
\begin{align}
	\gamma_\mu \gamma_\rho \gamma_\nu \gamma_5 &= \frac{1}{2} \left( \gamma_\mu \gamma_\rho \gamma_\nu \gamma_5 - \gamma_\nu \gamma_\rho \gamma_\mu \gamma_5 \right) \nonumber \\
	&= - \varepsilon_{\mu\rho\nu\alpha} \gamma_\alpha, \\
	\gamma_\mu \gamma_\nu \gamma_5 &= \frac{1}{2} \left(\gamma_\mu \gamma_\nu - \gamma_\nu \gamma_\mu \right) \gamma_5 = -i \sigma_{\mu\nu} \gamma_5, \\
	\gamma_\mu \gamma_\rho \gamma_\alpha \gamma_\nu \gamma_5 &= 2 i \epsilon_{\mu\rho\alpha\nu} \gamma_5 \gamma_5 = 2 i \varepsilon_{\mu\nu\rho\sigma}, 
\end{align}
where $\sigma_{\mu\nu} = \frac{i}{2} [\gamma_\mu, \gamma_\nu]$ and in writing the last line we have used, $\gamma_5 = \frac{i}{2} \varepsilon_{abcd} \gamma_a \gamma_b \gamma_c \gamma_d$.
Working out the numerator algebra and writing the amplitude in a more compact form,
\begin{align}
	\begin{split}
		T_{\mu\nu}^{VA}(k,q) &= \bar{u}(k,s^\prime) \left[ m^{(0)} + \mathcal{O}(m) \right] u(k,s) \\
		&+ (\mu \leftrightarrow \nu , q \to -q),
	\end{split}
\end{align}
where,
\begin{align}
	\begin{split}
		m^{(0)} &= \frac{1}{\tilde{Q}^2(k,q)} \Big[ i \varepsilon_{\mu\nu\rho\alpha} \gamma_\alpha \sin{(k+q)_\rho} \\
		&+ i \sigma_{\mu\nu} \gamma_5 \sum_\sigma (1 - \cos{(k+q)_\sigma}) \Big],
	\end{split}
\end{align}
we see that we pick up a term with tensor structure, which might be a concern. However, this tensor structure vanishes for a particle at rest and we are left with the simpler vector structure as we show below. From here on, we drop the $\mathcal{O}(m)$ term to simplify the notation.

Now we choose $\mu=1$, $\nu=3$, $\rho=2$, $\alpha=0$, and consider $\kb=(0,0,0)$ and $q_0=0$. Additionally, we use the following relations~\cite{Batelaan:2023jqp},
\begin{align}
	\begin{split} \label{eq:spivec}
		\bar{u}(\kb^\prime, \cdot) \gamma_0 u(\kb, \cdot) &= \left( s^\prime s + \frac{\kb^\prime \cdot \kb}{s^\prime s} \right) \mathbb{I} \\
		&+ \frac{i}{s^\prime s} (\kb^\prime \times \kb) \cdot \vec{\sigma},	
	\end{split}
	\\
	\begin{split} \label{eq:spitensor}
		\bar{u}(\kb^\prime, \cdot) \sigma_{ij} \gamma_5 u(\kb, \cdot) &= \varepsilon_{ijk} \left\{ \left( \frac{s^\prime}{s} \kb - \frac{s}{s^\prime} \kb^\prime \right) \mathbb{I} \right. \\
		&\left.+ i \left[ \left( \frac{s^\prime}{s} \kb + \frac{s}{s^\prime} \kb^\prime \right) \times \vec{\sigma} \right]_k \right\},
	\end{split}
\end{align}
where $s(\kb) = \sqrt{E(\kb) + m}$, and $(\vec{\sigma})_{\sigma^\prime \sigma} = \sigma \eb_3 \delta_{\sigma^\prime, \sigma} + (\eb_1 + i \sigma \eb_2) \delta_{\sigma^\prime, -\sigma}$ with $\sigma = \pm 1$. Then, for our choice of $\kb^\prime = \kb = (0,0,0)$, \Cref{eq:spivec,eq:spitensor} reduce to,
\begin{align}
	\bar{u}(\mathbf{0}, \cdot) \gamma_0 u(\mathbf{0}, \cdot) &= 2 \, m \, \mathbb{I}, \\
	\bar{u}(\mathbf{0}, \cdot) \sigma_{ij} \gamma_5 u(\mathbf{0}, \cdot) &= 0, 
\end{align}
which simplifies the amplitude,
\begin{equation} \label{eq:ca_latt}
	T_{13}^{VA}(k,q) = 4 \, i \, \varepsilon_{0123} \, m \, \frac{\sin{q_2}}{\tilde{Q}^2(q)},
\end{equation}
with,
\begin{equation}
	\tilde{Q}^2(\qb) = \sum_i \sin^2{\qb_i} + \left[ \sum_i (1 - \cos{\qb_i}) \right]^2,
\end{equation}
and the summation is understood to be running through spatial components.

\begin{figure}[t]
	\centering
	\includegraphics[width=.48\textwidth]{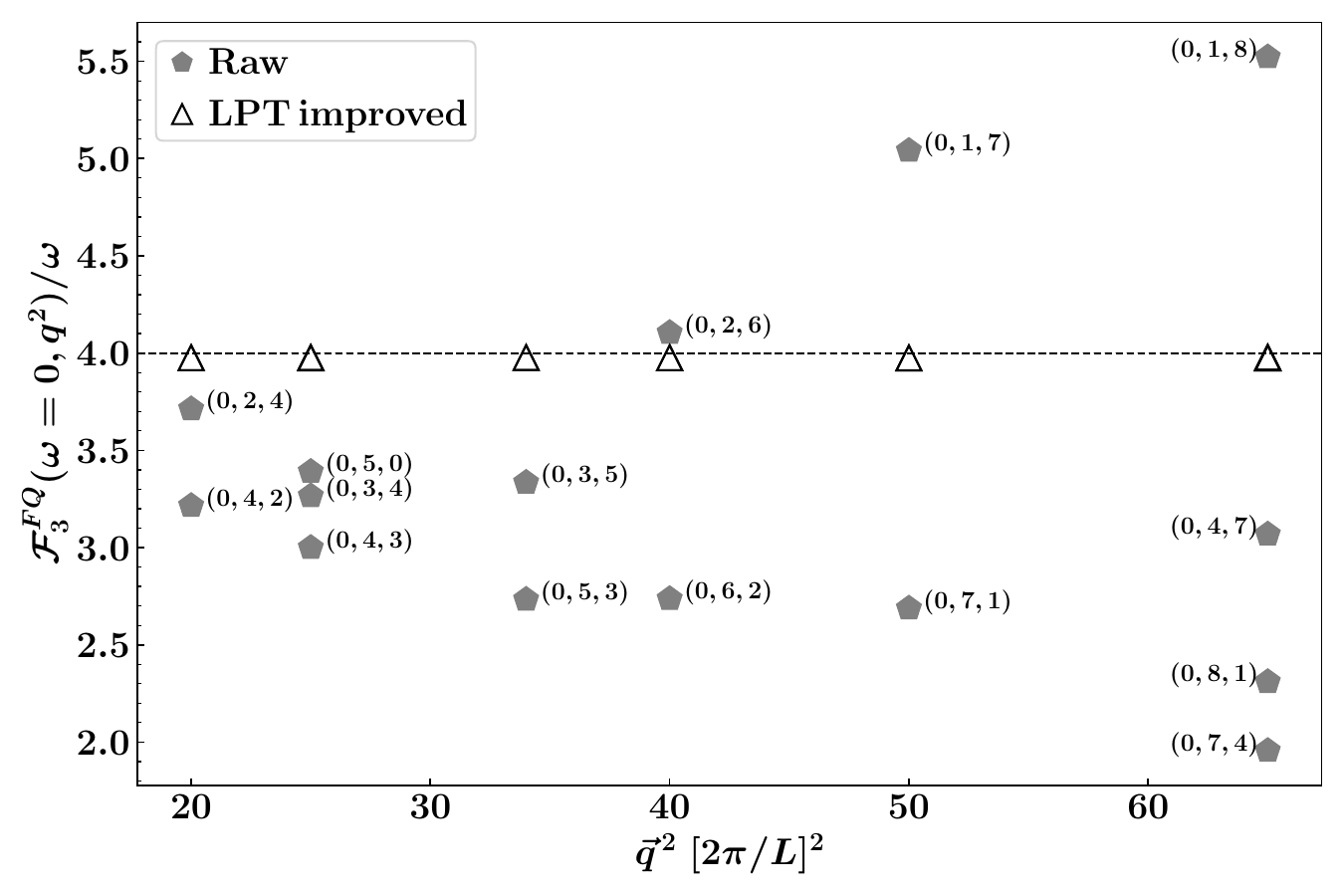}
	\caption{\label{fig:lpt_free_ca}Free field Compton structure function for some of the kinematics we consider in this work. The raw and LPT-improved points are denoted by filled pentagons and empty triangles, respectively. The LPT-improved points for the same $\qb^2$ coincide with each other. Dashed horizontal line denotes the expected continuum value.}
\end{figure}

In the continuum formulation, the amplitude for a structureless fermion is,
\begin{align} \label{eq:ca_cont}
	T_{13}^{VA}(k,q) &= i \, \varepsilon_{0123} \, m \, \frac{q_2}{Q^2} \frac{\mathcal{F}_3(\omega=0)}{\omega} \nonumber \\ 
	&= 4 \, i \, \varepsilon_{0123} \, m \, \frac{q_2}{Q^2},
\end{align}
where we have made use of the dispersion relation, \Cref{eq:disp}, in writing the second line. Finally, comparing \Cref{eq:ca_latt,eq:ca_cont} the correction becomes,
\begin{equation}
	\frac{q_2}{Q^2} \to \frac{\sin{q_2}}{\tilde{Q}^2(q)}.
\end{equation}
Since this correction is kinematical, it is universal, i.e. same for a hadron with internal structure, although higher-order corrections would require a calculation of relevant matrix elements.

We show the effect of the correction in \Cref{fig:lpt_free_ca} for a free (almost) massless quark. The raw points are the Compton structure function calculated on a free field gauge configuration (i.e. SU(3) gauge links are set to $\mathbb{I}$), $\mathcal{F}^{FQ}_3(\omega=0, q^2)$, following the analysis outlined in \Cref{sec:simu} (with a single-window plateau fitting instead of weighted averaging) at the almost-chiral limit, $\kappa = 0.1249$ ($a m_q = 0.003$). The LPT improvement shifts the results towards the expected continuum value and remedies the error due to broken rotational symmetry. 

\section{Elastic contribution to \texorpdfstring{$F_3$}{F3}}\label{app:f3_el}
The elastic contribution to the $F_3$ structure function is given by a combination of elastic form factors~\cite{Blunden:2011rd},
\begin{align}\label{eq:f3_el_app}
    F_{3,q}^{\rm \rm el.}(x,Q^2) = -G_A^{q}(Q^2) G_M^{q}(Q^2) \delta(1-x),
\end{align}
where $G_A^q$ and $G_M^q$ are the axial-vector and magnetic Sachs form factors, respectively, of a quark flavour $q$. Calculating the form factors follows the standard three-point function methodology. We summarise the main steps below for completeness.
\begin{figure*}[ht]
    \centering
    \includegraphics[width=0.85\linewidth]{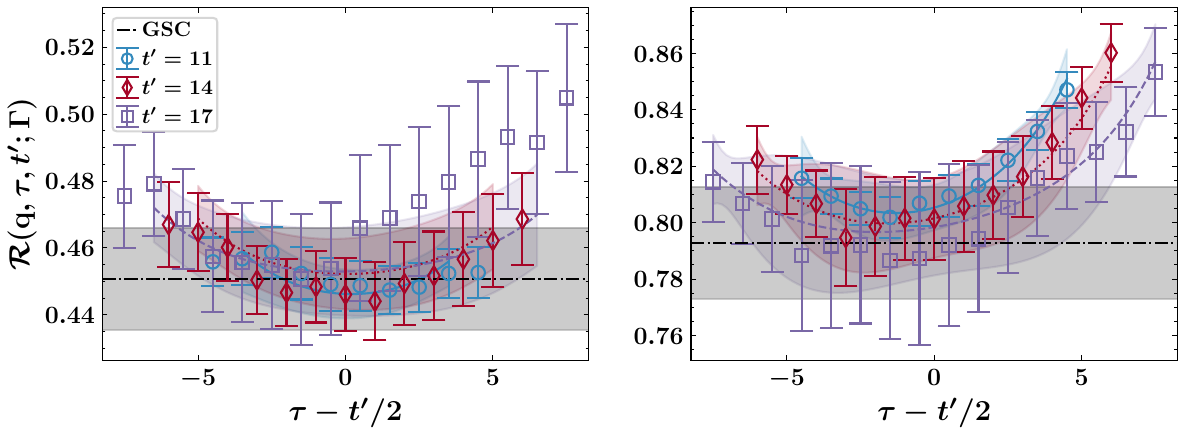}
    \caption{\label{fig:3pt} Euclidean time dependence of the ratio in Equation \eqref{eq:ratio_3pt} at three source-sink separations of $t^\prime=11$ (blue circles), $t^\prime=14$ (red diamonds) and $t^\prime=17$ (purple squares) for the operator $\overline{\psi}\gamma_2\psi$ (left) and the operator $\overline{\psi}\gamma_3\gamma_5\psi$ (right). Both plots make use of a fixed sink momentum of $\mathbf{p}'=(1,0,0)$, but different source momenta such that a non-zero signal is obtained, whilst maintaining the same $Q^2 = 0.14$ GeV$^2$. The left plot uses $\mathbf{p}=(1,0,1)$ and the spin-parity projector $\Gamma_1$, while the right plot uses $\mathbf{p}=(1,1,0)$ and the projector $\Gamma_3$. For these kinematics, the left (right) ratio is proportional to $G_M$ ($G_A$). The ground state contribution (GSC) is shown by the black band.}
\end{figure*}

\begin{figure*}[ht]
    \centering
    \includegraphics[width=.32\linewidth]{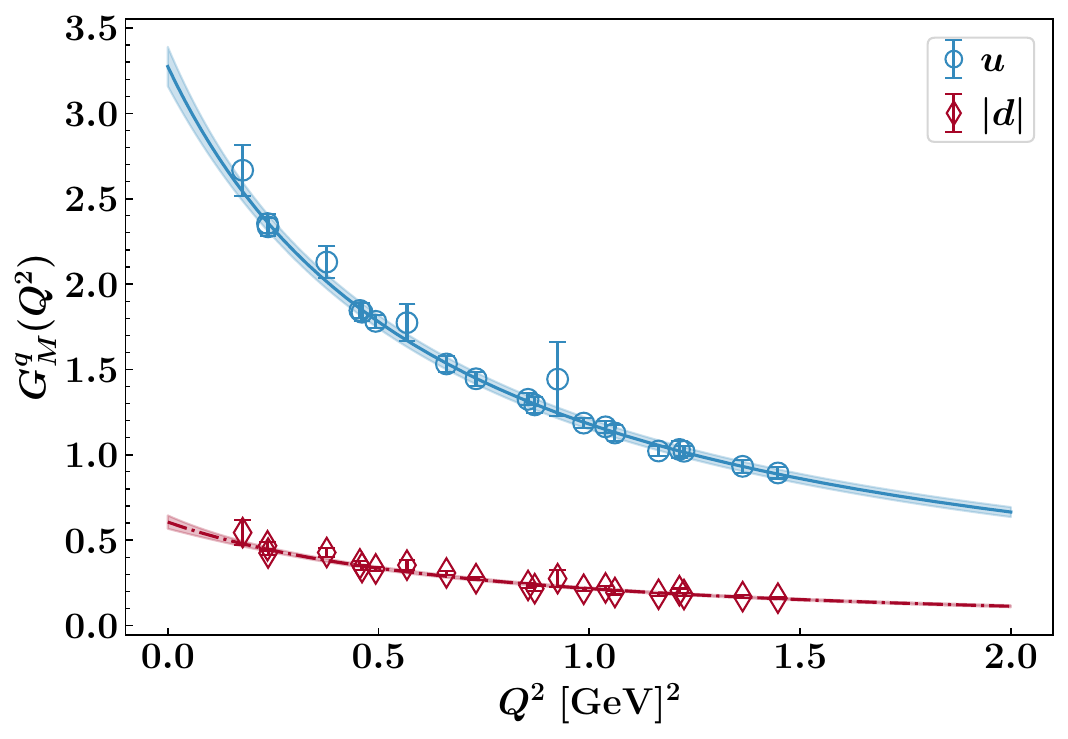}
    \includegraphics[width=.32\linewidth]{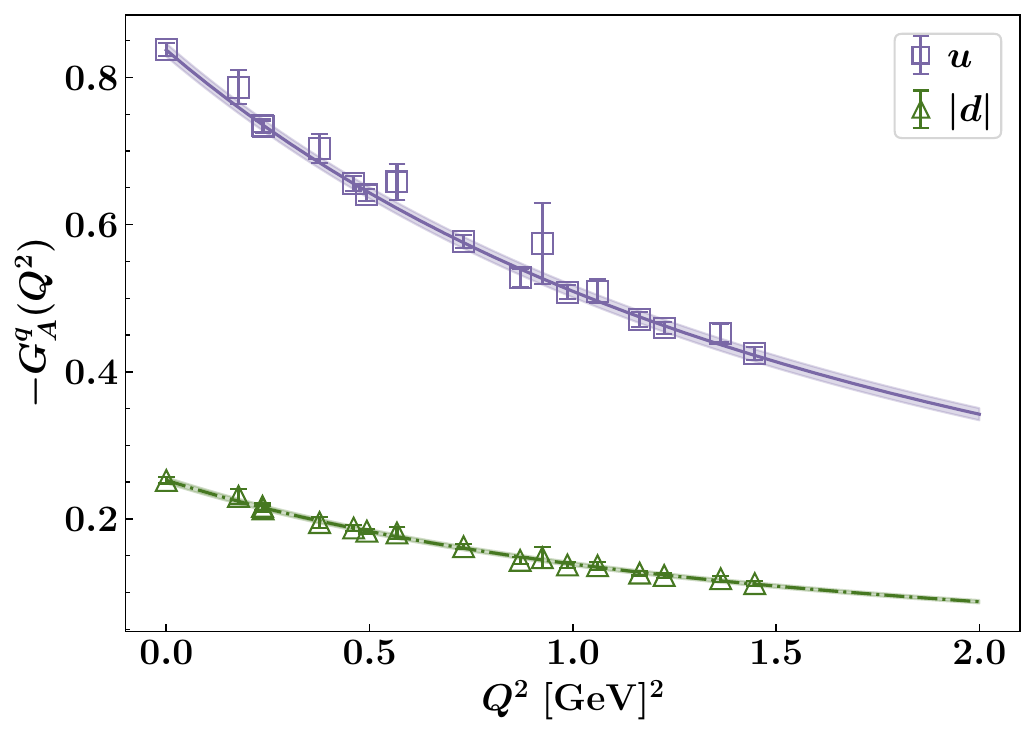}
    \includegraphics[width=.32\linewidth]{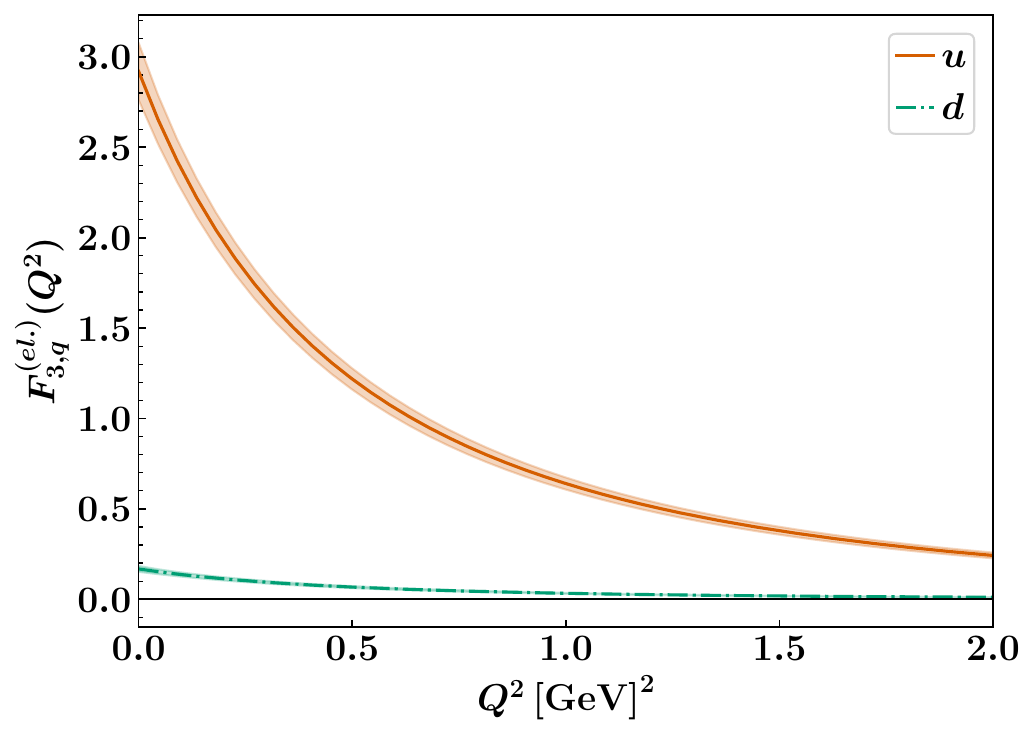}
    \caption{\label{fig:elastic} The $Q^2$ dependence of elastic form factors $G_M$ and $G_A$ along with the elastic contribution to the $F_3$ structure function (\Cref{eq:f3_el_app}). We show our results for the individual $u$- ($G_M$: blue circles, $G_A$: purple squares) and $d$-quark ($G_M$: red diamonds, $G_A$: green triangles) contributions obtained on the $\beta = 5.95$ ensemble considering the connected diagrams only. Note that the absolute value for the $d$-quark contribution is shown for $G_M$ and $G_A$ for clarity, and we plot $-G_A$ due to our choice of normalisation $G_A(0)=-g_A$. The $F_3^{\rm el.}$ is estimated using the parametric forms of $G_M$, \Cref{eq:pade} and $G_A$, \Cref{eq:dipole}.}
\end{figure*}

The matrix element of the electromagnetic current is decomposed in terms of Dirac ($F_1$) and Pauli ($F_2$) form factors. In Euclidean space it reads, 
\begin{align}
    \begin{split}
        \langle N(p^\prime, s^\prime) | J_\mu({\bf q}) | N(&p, s) \rangle = \bar u(p^\prime, s^\prime) \left[ \gamma_\mu F_1(Q^2) \right.\\ 
        &\left. + \, \sigma_{\mu\nu} \frac{q_\nu}{2 M_N} F_2(Q^2) \right] u(p,s),
    \end{split}
\end{align}
where $u(p,s)$ is a Dirac spinor with momentum $p$ and spin $s$, $q = p^\prime - p$, $Q^2 = -q^2$, and $\sigma_{\mu\nu} = i [\gamma_\mu, \gamma_\nu]/2$. The electromagnetic current is defined as,
\begin{equation}
    J_\mu = \frac{2}{3} \bar u \gamma_\mu u - \frac{1}{3} \bar d \gamma_\mu d + \cdots.
\end{equation}

The Sachs form factors are given by linear combinations of the Dirac and Pauli form factors,
\begin{align}
    G_E(Q^2) &= F_1(Q^2) - \frac{Q^2}{4 M_N^2} F_2(Q^2), \\
    G_M(Q^2) &= F_1(Q^2) +F_2(Q^2).
\end{align}

For the axial-vector current, the Lorentz decomposition of the matrix element is given by axial ($G_A$) and induced pseudo-scalar ($\tilde G_P$) form factors,
\begin{align}
    \begin{split}
        \langle N(p^\prime, s^\prime) | A_\mu({\bf q}) | N(p, s) \rangle &= \bar u(p^\prime, s^\prime) \left[ \gamma_\mu \gamma_5 G_A(Q^2) \right. \\ 
         &\left. + \gamma_5 \frac{q_\mu}{2 M_N} \tilde{G}_P(Q^2) \right] u(p, s),
    \end{split}
\end{align} 
where
\begin{equation}
    A_\mu = \bar u \gamma_\mu \gamma_5 u + \bar d \gamma_\mu \gamma_5 d + \cdots.
\end{equation}

The elastic form factors are calculated from ratios of three- and two-point functions. For the magnetic Sachs form factor $G_M$, we insert the local operators
\begin{equation}
    j_\mu = \overline{\psi}\gamma_\mu \psi, \hspace{0.5cm} \text{for } \mu = 1,2,3,4,
\end{equation}
and for the axial and induced pseudoscalar form factors, $G_A$ and $G_P$,
\begin{equation}
    j^A_\mu = \overline{\psi}\gamma_\mu\gamma_5\psi, \hspace{0.5cm} \text{for } \mu = 1,2,3,4.
\end{equation}
Matrix elements of these local operators are computed from three-point correlation functions,
\begin{align}
    \begin{split}
        G^{(3)}(\pb^\prime,t^\prime; \qb, \tau; \Gamma) &= \sum_{\mathbf{x}_1,\mathbf{x}_2} e^{-i\pb^\prime\cdot\mathbf{x}_2}e^{i\qb\cdot\mathbf{x}_1}\Gamma^{\alpha\beta} \times \\ 
        &\langle \chi_\beta(\mathbf{x}_2,t^\prime)\mathcal{O}(\mathbf{x}_1,\tau)\overline{\chi}_\alpha(\mathbf{0},0)\rangle,
    \end{split}
\end{align}
where $\chi$ are the standard nucleon interpolating operators. The local operator $\mathcal{O}$ is inserted at timeslice $\tau$ such that $t^\prime > \tau > 0$, where $t^\prime$ is the sink time. $\Gamma$ is a spin-parity projector, with $\Gamma_{\rm{unpol}}=\frac{1}{2}(I+\gamma_4)$ and $\Gamma_{i}=-\frac{i}{2}(I+\gamma_4)\gamma_i\gamma_5$, for $i=1,2,3$. We only consider connected contributions to the three-point functions in this analysis. We then construct the following ratio with the two-point functions,
\begin{align} \label{eq:ratio_3pt}
    \begin{split}
        &\mathcal{R}(\qb, \tau, t^\prime; \Gamma) = \frac{G^{(3)}(\pb^\prime,t^\prime; \mathbf{\qb}, \tau; \Gamma)}{G^{(2)}(\pb^\prime,t^\prime)} \times \\
        &\left[\frac{G^{(2)}(\pb^\prime,t^\prime)G^{(2)}(\pb^\prime,\tau)G^{(2)}(\pb,t^\prime-\tau)}{G^{(2)}(\pb,t^\prime)G^{(2)}(\pb,\tau)G^{(2)}(\pb^\prime,t^\prime-\tau)} \right]^{\frac{1}{2}},
    \end{split}
\end{align}
which, when assuming ground-state dominance at large Euclidean times, is proportional to the matrix element of the inserted operator. In order to control excited state contamination, we make use of a two-exponential ansatz for the three-point function when fitting directly to the ratio. The two-point functions are fit beforehand in order to extract the ground state energy $E_N$ and energy gap $\Delta E$, also using a two-exponential ansatz,
\begin{equation}
    G^{(2)}(\pb,t)=A_{2,0}e^{-E_N(\pb)t}(1 + A_{2,1} e^{-\Delta E(\pb) t}),
\end{equation}
where $A_{2,0},A_{2,1}$ are overlap factors. The three-point function ansatz is then
\begin{align}
    \begin{split}
        &G^{(3)}(\pb^\prime,t^\prime; \qb, \tau; \Gamma) = A_{3,0} e^{-E_N(\pb^\prime)(t^\prime-\tau)}e^{-E_N(\pb)\tau}\times \\
        & \left[ B_{00}+B_{10}e^{-\Delta E(\pb^\prime)(t^\prime-\tau)}+B_{01}e^{-\Delta E(\pb) \tau} + \right. \\
        & \left. B_{11}e^{-\Delta E(\pb)^\prime(t^\prime-\tau)}e^{-\Delta E(\pb)\tau)} \right],
    \end{split}
\end{align}
where $A_{3,0}$ is an overlap factor, and $B_{00}$ is proportional to the ground-state matrix element of interest. The terms $B_{01}$ and $B_{10}$ are proportional to transition matrix elements between the ground-state and first excited state (and vice versa), and the term $B_{11}$ is proportional to the matrix element between two nucleons in the first excited state. Representative plots for the ratio are shown in \Cref{fig:3pt} for $G_M$ and $G_A$ at a given $Q^2$.

Repeating the above analysis for a range of kinematics on our $\beta=5.65$ and $\beta=5.95$ ensembles, we determine the $Q^2$ dependence of the form factors. The Sachs magnetic form factor is modelled via a Pad\'e approximant inspired by Kelly's parametrisation~\cite{Kelly:2004hm},
\begin{equation} \label{eq:pade}
    G_M(Q^2) = \frac{a_0}{1 + b_1 \tau + b_2 \tau^2},
\end{equation}
where $\tau = Q^2/(2 M_N)^2$, and $a_0$, $b_{1,2}$ are free fit parameters. We assume a dipole form for the axial form factor,
\begin{equation} \label{eq:dipole}
    G_A(Q^2) = \frac{G_A(0)}{(1 + Q^2/\Lambda^2)^2},
\end{equation} 
with $G_A(0) = -g_A$ the axial charge, following PDG definition~\cite{ParticleDataGroup:2024cfk} and $\Lambda$ the dipole mass. We fix $G_A(0)$ to its respective (ensemble and flavour) values calculated in Ref.~\cite{Smail:2023qlx} and determine the dipole mass in this analysis. Our results for $G_M$, $G_A$, and $F_3^{\rm el.}$ estimateted via \Cref{eq:f3_el_app} are shown in \Cref{fig:elastic} for a representative case.

\end{document}